\documentclass[conference]{IEEEtran}

\usepackage{amssymb}
\usepackage{amsmath}
\usepackage{bm}
\usepackage{graphicx}
\usepackage{xcolor}
\usepackage{import}
\usepackage[utf8]{inputenc}
\usepackage{listings}
\usepackage[noend]{algpseudocode}
\usepackage[hyphens]{url}
\usepackage{multirow}
\usepackage[font={small,it}]{caption}
\usepackage[bookmarks=true]{hyperref} 
\usepackage{array}
\usepackage{rotating}   
\usepackage{makecell}
\usepackage{colortbl}
\usepackage{enumerate}
\usepackage{booktabs}
\usepackage{subfig}
\usepackage{paralist}
\usepackage[ruled,vlined]{algorithm2e}
\algrenewcommand\algorithmiccomment[1]{\hfill \textcolor{gray}{$\triangleright$ \textit{#1}}}

\DeclareUnicodeCharacter{00A0}{ }

\newcommand{\varDash}[1]{{\operatorname{\mathit{#1}}}}

\newcommand{\specialcell}[2][c]{
	\begin{tabular}[#1]{@{}l@{}}#2\end{tabular}}

\renewcommand{\paragraph}[1]{\vspace{0.1cm}\noindent{\bf #1.}}

\renewcommand{\tablename}{Table}

\renewcommand{\figurename}{Figure}

\begin{document}
\title{ASNM Datasets: A Collection of Network Traffic Features for Testing of Adversarial Classifiers and Network Intrusion Detectors}

\author{
	\IEEEauthorblockN{
		Ivan Homoliak\IEEEauthorrefmark{1} \\
		ihomoliak@fit.vutbr.cz
	}
	\and
	\IEEEauthorblockN{
		Petr Hanacek\IEEEauthorrefmark{1}\\
		hanacek@fit.vutbr.cz
	}
	\and

	\and
	\IEEEauthorblockA{
		\hspace{3.5cm}
		\IEEEauthorrefmark{1}Faculty of Information Technology, Brno University of Technology
	}
}

\maketitle

\begin{abstract}
    In this paper, we present three datasets that have been built from network traffic traces using ASNM features, designed in our previous work.
	The first dataset was built using a state-of-the-art dataset called CDX 2009, while the remaining two datasets were collected by us in 2015 and 2018, respectively.
	These two datasets contain several adversarial obfuscation techniques that were applied onto malicious as well as legitimate traffic 
	samples during ``the execution'' of particular TCP network connections.
	Adversarial obfuscation techniques were used for evading machine learning-based network intrusion detection classifiers.
	Further, we showed that the performance of such classifiers can be improved when partially augmenting their training data by samples obtained from obfuscation techniques.
	In detail, we utilized tunneling obfuscation in HTTP(S) protocol and non-payload-based obfuscations modifying various properties of network traffic by, e.g., TCP segmentation, re-transmissions, corrupting and reordering of packets, etc. 
	To the best of our knowledge, this is the first collection of network traffic metadata that contains adversarial techniques and is intended for non-payload-based network intrusion detection and adversarial classification.
	Provided datasets enable testing of the evasion resistance of arbitrary classifier that is using ASNM features. 
\end{abstract}

\section{Introduction}
Network intrusion attacks such as exploiting unpatched services are one of the most dangerous threats in the domain of information security~\cite{mcafeeIntrusions},~\cite{certTargetVulns2015}.
Due to an increasing sophistication in the techniques used by attackers, misuse-based/knowledge-based~\cite{debar2000revised} intrusion detection suffers from undetected attacks such as zero-day attacks or polymorphism, enabling an exploit-code to avoid positive signature matching of the packet payload data. 
Therefore, researchers and developers are motivated to design new methods to detect various versions of the modified network attacks including the zero-day ones.
These goals motivate the popularity of Anomaly Detection Systems (ADS)
and also the classification-based approaches in the context of intrusion detection. 
Anomaly-based approaches are based on building profiles of normal users, and they try to detect anomalies deviating from these profiles~\cite{debar2000revised}, which might lead to detection of unknown intrusions, but on the other hand it might also generate many false positives.
In contrast, the classification-based approaches take advantage of both misuse-based and anomaly-based models in order to leverage their respective advantages. 
The classification-based detection methods first build a model based on the labeled samples from both classes -- intrusions and the legitimate instances.
Second, they compare a new input to the model and select the more similar class as the predicted label.
Classification and anomaly-based approaches are capable to detect some unknown intrusions, but at the same time they may be susceptible to evasion by obfuscation techniques.

\medskip
In this paper, we present ASNM datasets, a collection of malicious and benign network traffic data. 
ASNM datasets include records consisting of several features that express miscellaneous properties and characteristics of TCP communications (i.e., aggregated bidirectional flows). 
These features are called Advanced Security Network Metrics (ASNM) and were designed in our previous work~\cite{homoliakasnm2013} with the intention to distinguish between legitimate and malicious TCP connections (i.e., intrusions and C\&C channels of malware).
ASNM features are extracted from tcpdump~\cite{tcpdump} traces and do not perform deep packet inspection during their computation, which makes them suitable for passive monitoring of (potentially encrypted) network traffic.

To this end, we performed ASNM feature extraction over three different subsets of network traffic collections, resulting in three sub-datasets that we provide to the community:
\begin{itemize}
	\item \textbf{ASNM-CDX-2009 Dataset}: was created from tcpdump traces of CDX 2009 dataset~\cite{sangster2009toward}. 
	The dataset misses a few newer ASNM features and does not contain any obfuscations of the network traffic (see details in \autoref{sec:desc-CDX}).
	
	\item \textbf{ASNM-TUN Dataset:} was created with the intention to evade and improve machine learning classifiers, and besides legitimate network traffic samples, it contains tunneling obfuscation technique~\cite{homoliak:NBAofObfNetVul} applied onto malicious network traffic, in which several vulnerable network services were exploited (see details in \autoref{sec:desc-TUN}).
	
	\item \textbf{ASNM-NPBO Dataset:} like the previous dataset, the current dataset was created with the intention to evade and improve machine learning classifiers, and it contains non-payload-based obfuscation techniques (modifying the properties of network flows) applied onto malicious traffic and onto several samples of legitimate traffic (see details in \autoref{sec:desc-NPBO}).
	
\end{itemize}
All ASNM datasets are available for download at~\href{http://www.fit.vutbr.cz/~ihomoliak/asnm/}{http://www.fit.vutbr.cz/$\sim$ihomoliak/asnm/}.
In the following, we will describe ASNM features, detail particular datasets, and finally benchmark several supervised classification methods used for non-payload-based\footnote{Not performing deep packet inspection.} network intrusion detection in ASNM datasets.
We conduct a few experiments aimed at the adversarial classification, and we demonstrate that proposed obfuscations are able to evade an intrusion detection of employed classifiers.
Consequently, we show that after partially augmenting the training data by obfuscated attacks, we can significantly improve the performance of the classifiers.

The rest of the paper is organized as follows.
In \autoref{sec:background}, we define the classification problem in intrusion detection and describe preliminaries and terms used throughout the paper.
In \autoref{sec:asnm-features}, we formally define ASNM features and describe them.
Next, we introduce particular ASNM datasets in \autoref{sec:ASNM-datasets} and consequently perform their benchmarking in \autoref{sec:benchmarking}.
In \autoref{sec:discussion} we discuss limitations of the proposed datasets.
Then, in \autoref{sec:related-work}, we mention existing network datasets and network features, compare them to the ASNM datasets, and finally in \autoref{sec:conclusion} we conclude the paper. 
\vspace{0.3cm}
\section{Problem Definition and Preliminaries}\label{sec:background}
First, we define the scope of our work by introducing the network connection as an elementary data object that is used for building our datasets.
Second, we describe the feature extraction process over a network connection object, which forms a sample/data record in our datasets.
Then, we describe the intrusion detection classification task, representing the problem that is addressed by an arbitrary binary classifier given a dataset containing 2-class labels.
This problem represents the main challenge of ASNM datasets, but the application of ASNM datasets can be straightforwardly extended to a multi-class classification problem in sub-datasets containing multi-class labels.

\subsection{TCP Connection}\label{sec:tcp-con}
Consider a session of a protocol at the application layer of the TCP/IP stack that serves for data transfer between the client/server based application.
The interpretation of application data exchanges between client and server can be formulated, considering the TCP/IP stack up to the transport layer, by connection $c$ 
that is constrained to the connection-oriented protocol TCP at L4, Internet protocol IP at L3, and Ethernet protocol at L2.
The TCP connection $c$ is represented by the tuple
\[
c = (t_{s}, t_{e}, p_{c}, p_{s}, ip_{c}, ip_{s}, P_{c}, P_{s}),
\]
which consists of the start and end timestamps $t_s$ and $t_e$, ports of the client and the server $p_c$ and $p_s$, IP addresses of the client and the server  $ip_c$ and $ip_s$, sets of packets sent by the client $P_{c},$ and by the server $P_{s}$, respectively (see details in \autoref{tab:TCP_connection_tuple_items} of Appendix).
Sets $P_{c}$ and $P_{s}$ contain a number of packets, where each of them can be interpreted by the packet tuple 
\[
p = (t, size, eth_{src}, eth_{dst}, ip_{off}, ip_{ttl}, ip_{p}, ip_{sum},
\]
\vspace*{-0.5cm}
\[
ip_{src}, ip_{dst}, ip_{dscp}, tcp_{sport}, tcp_{dport}, tcp_{sum},
\]
\vspace*{-0.5cm}
\[
tcp_{seq}, tcp_{ack}, tcp_{off}, tcp_{flags}, tcp_{win}, tcp_{urp}, data).
\]
The symbols of the tuple are described in \autoref{tab:Packet_tuple_items} of Appendix.
We assume that the payload of $P_{s}$ and $P_{c}$ is encrypted, and thus data of these packet sets are not accessible.

\medskip
Each TCP connection has its beginning that is represented by a~three way handshake, in which, three packets that contain the same IP addresses ($ip_{s}$, $ip_{d}$), ports ($p_{s}$, $p_{d}$), and sequence/acknowledgment numbers $(tcp_{seq}$, $tcp_{ack})$ conforming the specification of RFC 793\footnote{URL http://www.ietf.org/rfc/rfc793.txt, page 30.} must be found.
Similarly, each TCP connection has its end, which is defined by a three-way-endshake or by an inactivity timeout.\footnote{E.g., in Unix-based systems, such a timeout is equal to five days.}

\subsection{Feature Extraction}\label{sec:feature-extraction}
At this time, we can express characteristics of a TCP connection by network connection features. 
The features extraction process is defined as a function that maps a connection $c$ into space of features $F$: 
\begin{eqnarray}
\begin{split}           
f(c) &\mapsto F, \\
F = (F_1, &~F_2, \ldots, F_n), \\
\end{split}
\end{eqnarray}
where $n$ represents the number of defined features.
Each function $f_i$ that extracts feature $i$ is defined as a mapping of a connection $c$ into feature space~$F_i$: 
\begin{eqnarray}
f_i(c) \mapsto F_i, ~~i \in \{1,\ldots,n\},      
\end{eqnarray}
and each element\footnote{Representing a particular dimension of a feature.} of codomain $F_i$ is defined as 
\begin{eqnarray}
\label{featureElementDefinition}
\begin{split}
e = (e_0,\ldots,&e_n), ~n \in \mathbb{N}_0,\\
e_i \in \mathbb{N} ~\mid ~e_i \in \mathbb{R} ~\mid e_i \in & ~\Gamma^{+}, ~i \in \{0,\ldots,n\},\\
\Gamma = \{a-z, A&-Z, 0-9\},
\end{split}
\end{eqnarray}
where $\Gamma^{+}$ denotes positive iteration of the set $\Gamma$.
Note that for demonstration purposes, we abstract in our formalization from the fact that some features of a network connection $c$ can be extracted not only from $c$ itself but in addition from metadata of $c$ that are not part of $c$. 
For example, such metadata may represent ``neighboring'' network connections of $c$, which we later refer to as \textit{a context} of $c$ (see \autoref{sec:asnm-features}).

In general, network connection features can be instantiated, for example, by discriminators of A. Moore~\cite{moore2005discriminators}, Kyoto 2006+ features~\cite{song2011statistical}, basic and traffic features\footnote{Not content features, which work over payload of the network data.} of KDD Cup'99 dataset~\cite{KDDCup99web}, NetFlow features~\cite{netflow}, or ASNM features~\cite{homoliakasnm2013}, CICFlowMeter features~\cite{lashkari2017characterization}, multi-layered network traffic features from BGU~\cite{bekerman2015unknown}, or connection-less features~\cite{homoliak2016features}.

\subsection{Intrusion Detection Classification Task}\label{ADSPredictionTask}
A data sample of the dataset $D_{tr}$ refers to the vector of the network connection features, defined in \autoref{sec:feature-extraction}.
Then, referring to~\cite{kohavi1995study}, let $X = V \times Y$ be the space of labeled samples, where $V$~represents the space of unlabeled samples and $Y$ represents the space of possible labels.
Let $D_{tr} = \{x_1, x_2, \ldots, x_n\}$ be a training dataset consisting of $n$~labeled samples, where 
\begin{eqnarray}
x_i = (v_i \in V,~y_i \in Y).
\end{eqnarray}
Consider classifier $C$ which maps unlabeled sample $v \in V$ to a label $y \in Y$: 
\begin{eqnarray}
y = C(v),
\end{eqnarray}
and learning algorithm $A$ which maps the given dataset $D$ to a classifier $C$: 
\begin{eqnarray}
C = A(D_{tr}).
\end{eqnarray}
The notation $y_{predict} = A(D_{tr}, v)$ denotes the label assigned to an unlabeled sample $v$ by the classifier $C$, build by learning algorithm $A$ on the dataset $D_{tr}$.
Then, all extracted features $f()$ of an unknown connection $c$ can be used as an input of the trained classifier $C$ that predicts the target label:
\begin{eqnarray}
y_{predict} = A\big(D_{tr}, f(c)\big),
\end{eqnarray}
where 
\begin{eqnarray}
y_{predict} \in \{\mathit{Intrusion}, \mathit{Legitimate}\}.
\end{eqnarray}

\subsection{Adversarial Obfuscations \& Evasion of the Classifier}\label{sec:adversarial-obfuscations}
Assume a connection $c_{m}$ representing a malicious communication
executed without any obfuscation. Then, $c_{m}$ can be expressed
by network connection features
\begin{eqnarray}
f(c_{m}) \mapsto F^m = (F_1^m, &F^m_2, \ldots, F^m_n)
\end{eqnarray} that are delivered to the previously trained classifier $C$. Assume
that $C$ can correctly predict the target label as a malicious one,
because its knowledge base is derived from training dataset $D_{tr}$
containing features of malicious connections having similar (or the same) behavioral characteristics as~$c_m$. 

Now, consider connection $c_{m}'$ that represents the malicious communication $c_{m}$ executed by employment of an obfuscation technique that is
aimed at modification of network behavioral properties of the connection $c_m$. 
An obfuscation technique
can modify $P_{c}$ and $P_{s}$ packet sets of the original connection
$c_{m}$ as well as IP addresses ($ip_{s}$, $ip_{d}$) and ports ($p_{s}$, $p_{d}$) of the original connection $c_m$.

Hence, network connection features extracted for $c_{m}'$ are represented
by 
\begin{eqnarray}
f(c_{m}') \mapsto F^{m'} = (F_1^{m'}, &F^{m'}_2, \ldots, F^{m'}_n)
\end{eqnarray} and have different values than features $F^{m}$ of the connection
$c_{m}$. 
Therefore, we conjecture that the likelihood of a correct prediction of
of $c_{m}'$-connection's features $F^{m'}$ by the previously assumed
classifier $C$ is lower than in the case of connection $c_{m}$, which might cause an evasion of the detection.
Also, we conjecture that the classifier $C'$ trained by learning algorithm
$A$ on training dataset $D_{tr}'$, containing obfuscated malicious
instances, will be able to correctly predict higher number of unknown
obfuscated malicious connections than classifier $C$. 
We will demonstrate the correctness of these assumptions in \autoref{sec:benchmarking} on two of our datasets. 
\vspace{0.3cm}
\section{ASNM Features and Context Analysis}\label{sec:asnm-features}
ASNM features~\cite{homoliakasnm2013} are network connection features that describe various properties of TCP connections and were designed with the intention to distinguish between legitimate traffic and remote buffer overflow attacks.\footnote{See Appendix D of~\cite{2016-ihomoliak-thesis} for the full list of ASNM features.} 
We studied behavioral characteristics of remote buffer overflow attacks in our previous work~\cite{marovs2013detection}, and our findings inspired the design of ASNM features. 
We can interpret ASNM features like an extended protocol NetFlow~\cite{netflow} but describing more than statistical properties of network connections. 
In addition to NetFlow features, ASNM features represent dynamical, localization, and, most importantly, the behavioral properties of network connections.
Moreover, some of the features utilize a context of an analyzed connection $c$, which represents ``neighboring'' connection objects (see \autoref{sec:context}).

In the following, we assume an input dataset of network traffic traces, which is used for identification of network TCP connection objects $C=\{c_1,\ldots,c_M\}$, where $M$ is a count of TCP connections in the dataset.

\subsection{Context Definition}\label{sec:context}
We assume a dataset of TCP connection objects (as described in \autoref{sec:background})
Considering analyzed TCP connection $c_k$, we define a~sliding window ${sw}$ of length $\tau$ as a~set of TCP connections $W_{k}$ that
are delimited by $\pm \, \frac{\tau}{2}$:
\begin{eqnarray}
\begin{split}
{sw}({c}_k, \tau) &= W_{k}\\
W_k &\subseteq C,\\
W_{k} &= \{{c}_{j}\},
\end{split}
\end{eqnarray}
where each TCP connection ${c}_j$ must satisfy the following:        
\begin{eqnarray}
\begin{split}        
{c}_{j}[t_{s}] >{c}_{k}[t_{s}]-\frac{\operatorname{\tau}}{2},\\
{c}_{j}[t_{e}] <{c}_{k}[t_{s}]+\frac{\operatorname{\tau}}{2}.
\end{split}
\end{eqnarray}
The next fact about each particular TCP connection ${c}_{k}$
is an unambiguous association of it to particular sliding
window $W_{k}$. We can interpret the start time $t_{s}$ of the TCP
connection ${c}_{k}$ as a~center of the sliding window $W_{k}$.
Then, we can denote a~shift of the sliding window $\Delta(W_{j})$
which is defined by start time differences of two consecutive TCP
connections in $C$:

\begin{eqnarray}
\begin{split}\Delta(W_{j})={c}_{j+1} & [t_{s}]-{c}_{j}[t_{s}],\\
j\in\{1,\ldots & ,|C|-1\}.
\end{split}
\end{eqnarray}
Next, we define the context $K_k$ of the TCP connection $c_k$, which is a~set of
all connections in a~particular sliding window $W_{k}$ excluding
analyzed TCP connection ${c}_{k}$:

\begin{eqnarray}
K_{k}=\{{c}_{1},\ldots,{c}_{n}\}=\{W_{k}~\backslash~{c}_{k}\}.
\end{eqnarray}

\begin{figure}
	\begin{centering}
		\includegraphics[scale=0.48]{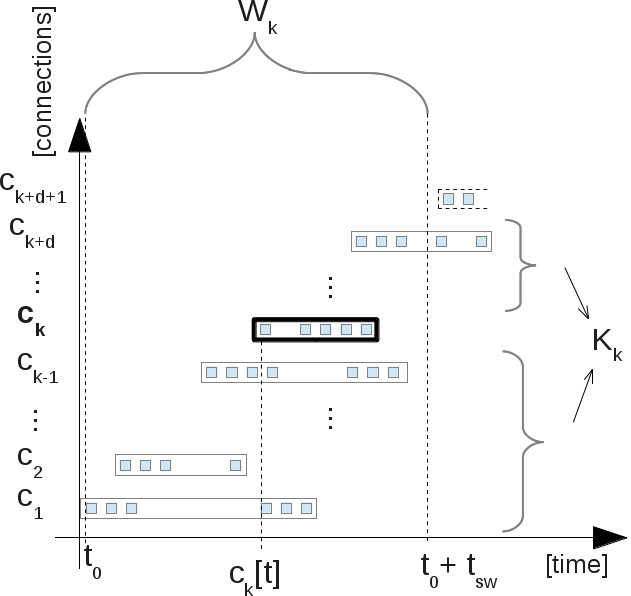}
		\par\end{centering}
	
	\caption{Sliding window and the context of the connection ${c}_{k}$
		\cite{homoliakasnm2013}.}\label{fig:Sliding-window-fig}
\end{figure}

Defined terms are shown in \autoref{fig:Sliding-window-fig}. 
In the figure, the $x$ axis displays time, and the $y$ axis represents TCP
connections, which are shown in the order of their occurrences. Packets are
represented by small squares, and TCP connections are represented by
a rectangular boundary of particular packets. A bold line and bold font
are used for depicting an analyzed TCP connection ${c}_{k}$, which
has an associated sliding window $W_{k}$ and context $K_{k}$. TCP
connections, which are part of the sliding window $W_{k}$, are drawn
by full line boundary, and TCP connections, which are not part of this
sliding window, are drawn by a~dashed line boundary.
We note that only a few features from the ASNM datasets utilize context; these features belong to dynamic and behavioral categories (see \autoref{sec:categories-of-asnm-features}).

\subsection{ASNM Feature Extraction}
In addition to a general definition of network connection feature extraction (see \autoref{sec:feature-extraction}), we incorporate the context of a TCP connection into the extraction process of ASNM features.
The ASNM feature extraction is thus defined as a~function that maps a~connection ${c}_k$ with its context $K_k = {sw}({c}_k, \tau)$ into feature space $F$:
\begin{eqnarray}
\begin{split}           
f({c}_k, K_k) &\mapsto F, \\
F = (F_1, F_2, &\ldots, F_n), \\
\end{split}
\end{eqnarray}
where $n$ represents the number of defined features, while the rest of the definition is inherited from \autoref{sec:feature-extraction}.

\subsection{Categorization of ASNM Features}\label{sec:categories-of-asnm-features}
The list of original proposed ASNM feature set contains $167$ features and is present Master's thesis~\cite{DP_metriky_ih} and formally described in~\cite{homoliakasnm2013}.
However, the ASNM feature set was later extended~\cite{2016-ihomoliak-thesis}, resulting in 194 features.
These $194$ features are in many cases a~result of reasonable parametrization of the base feature functions $f_i()$, 
We depict a categorization of our feature set in \autoref{tab:Distribution-of-Metrics} together with their counts. 
We decided to determine the naming of particular categories of features according to their principles, not according to their data representation.
In the following, we briefly describe each category. 
\begin{table}
	\begin{centering}
		\begin{tabular*}{0.2\textwidth}{@{\extracolsep{\fill} }r c}         \toprule         \textbf{\specialcell{~~~~~Category of\\ASNM Features}} & $\mathbf{\#}$  \\ \midrule
			Statistical  & 77\\
			Dynamic      & 32\\
			Localization & 8\\
			Distributed  & 34\\
			Behavioral   & 43\\
			\bottomrule
		\end{tabular*}
		\par\end{centering}
	
	\caption{Categorization of ASNM features.}
	\label{tab:Distribution-of-Metrics}
\end{table}

\paragraph{Statistical Features}
In this category of ASNM features, the statistical properties of TCP
connections are identified. All packets of the TCP connection are
considered in order to determine count, mode, median, mean, standard
deviation, ratios of some header fields of packets, or the packets
themselves. This category of features partially uses a~time representation
of packets occurrences, in contrast to the dynamic category (see below).
Therefore, it includes particularly dynamic properties of the analyzed
TCP connection, but without any context. Most of the features
in this category also distinguish inbound and outbound packets of
the analyzed TCP connection.

\paragraph{Dynamic Features}
Dynamic features were defined with the aim to examine dynamic properties
of the analyzed TCP connection and transfer channel such as a speed
or an error rate. These properties can be real or simulated. Eighteen
of the features consider the context of an analyzed TCP connection.
The difference between some of the statistical and dynamic features
from a~dynamic view can be demonstrated on two instances of the same
TCP connection, which performs the same packet transfers, but in different
context conditions and with different packet retransmission and attempts
to start or finish the TCP connection. 
Many of the defined features distinguish between the inbound
and outbound direction of the packets and consider the statistical properties
of the packets and their sizes, as mentioned in statistical features.

\paragraph{Localization Features}
The main characteristic of the localization features category is that
it contains static properties of the TCP connection. These properties
represent the localization of participating machines and their ports
used for communication. In some features, the localization is expressed
indirectly by a~flag, which distinguishes whether participating machines
lay in a~local network or not. Features included in this category
do not consider the context of the analyzed TCP connection, but they
distinguish a~direction of the analyzed TCP connection.

\paragraph{Distributed Features}
The characteristic property of the distributed features category is the
fact that they distribute packets or their lengths to a~fixed number
of intervals per the unit time specified by a~logarithmic scale (1s, 4s,
8s, 32s, 64s). A logarithmic scale of fixed time intervals was proposed
as a~performance optimization during the extraction of the features. The
next principal property of this category is vector representation.
All these features are supposed to work within the context of an analyzed
TCP connection.

\paragraph{Behavioral Features}
Behavioral features represent properties associated with the behavior of an analyzed TCP connection.
Examples include legal or illegal connection closing, the polynomial approximation of packet lengths in a~time domain or an index of occurrence domain, count of new TCP connections after starting an analyzed TCP connection, coefficients of Fourier series with the distinguished direction of an analyzed TCP connection, etc. 
\vspace{0.3cm}
\section{ASNM Datasets}\label{sec:ASNM-datasets}
In this section, we detail three different datasets that have been built using ASNM features.
The first of them was built using an existing dataset of network traffic traces, while the remaining two were collected by us, and they contain several adversarial obfuscation techniques that were applied onto malicious as well as legitimate samples during ``the execution'' of particular network connections.

\subsection{ASNM-CDX-2009 Dataset}\label{sec:desc-CDX}
ASNM-CDX-2009 dataset was build from CDX-2009 dataset~\cite{CDX-dataset-web}, which was introduced by Sangster et al.~\cite{sangster2009toward} and it contains data in tcpdump format as well as SNORT~\cite{snortWeb} intrusion prevention logs, as relevant sources for our purpose. 

The CDX 2009 dataset was created during the network warfare competition, in which one of the goals was to generate a labeled dataset. 
By labeled dataset, the authors mean tcpdump traces of all simulated communications and SNORT log with information about occurrences of intrusions, deemed as the expert knowledge. 
Network infrastructure contained four servers with four vulnerable services (one per each server), while the authors provided two collections of network traces: 1) network traces captured outside the West Point network border and 2) network traces captured by National Security Agency (NSA).
The services that run on the hosted servers together with IP addresses of the servers are listed in \autoref{CDX_2009_servers}. 
\begin{table}[h]
	\centering
	\begin{tabular}{r l l l} 
		\toprule
		\textbf{Service} & \textbf{OS} & \textbf{Internal IP} & \textbf{External IP} \\ \midrule
		\textbf{Postfix Email} & FreeBSD & 7.204.241.161 & 10.1.60.25 \\
		\textbf{Apache Web Server} & Fedora 10 & 154.241.88.201 & 10.1.60.187\\ 
		\textbf{OpenFire Chat} & FreeBSD & 180.242.137.181 & 10.1.60.73\\
		\textbf{BIND DNS} & FreeBSD & 65.190.233.37 & 10.1.60.5\\              
		\bottomrule
	\end{tabular}
	\caption{A list of vulnerable servers in CDX 2009 dataset.}    
	\label{CDX_2009_servers}
\end{table}
Two types of IP addresses are shown in this table: 
\begin{itemize}
	\item \textbf{Internal IP} addresses -- corresponding to the SNORT log,
	\item \textbf{External IP} addresses -- corresponding to a TCP dump network captured outside the West Point network border.
\end{itemize}
Note that specific versions of services described in \cite{sangster2009toward} were not announced.
We found out that SNORT log can be associated only with data capture outside of West point network border and only with significant timestamps differences -- approximately $930$ days. 
We have not found any association between SNORT log and data capture performed by NSA. 
We focused only on buffer overflow attacks found in SNORT log, and we performed a match with the packets contained in the West point network border capture. 

Despite all the efforts, we matched only $44$ buffer overflow attacks out of $65$ entries in SNORT log.
To correctly match SNORT entries, it was necessary to remap external IP addresses to internal ones, because SNORT detection was performed in external network and TCP dump data capture contains entries with internal IP addresses.
We found out that in CDX 2009 dataset, buffer overflow attacks were performed only on two services -- Postfix Email and Apache Web Server.

The buffer overflow attacks that were matched with data capture have their content only in two TCP dump files: 
\begin{itemize}
	\item \begin{em}2009-04-21-07-47-35.dmp\end{em}
	\item \begin{em}2009-04-21-07-47-35.dmp2\end{em}
\end{itemize}
Due to the high count of all packets (approx. 4 mil.) in all dump files, we decided to consider only these two files for the purpose of extraction both malicious and legitimate samples (which together contain $1,538,182$ packets). 
We also noticed that network data density was increased in the time when attacks were performed. 
Consequently, we made another reduction of all packets consider so far, which filtered enough temporal neighborhood of all attacks occurrences, and at the same time, included a high enough number of legitimate TCP connections. 
In the result, we used $204~953$ packets for the extraction of ASNM features.
\begin{table}
	\centering    
	
	\begin{tabular}{>{\raggedleft}p{2.1cm}c>{\centering}p{2.0cm}>{\centering}p{1.8cm}>{\centering}p{1.8cm}}
		\toprule
		\multirow{2}{3.6cm}{\textbf{~Network Service}} & \multicolumn{3}{c}{\textbf{Count of TCP Connections}} \smallskip \tabularnewline
		\expandafter\cline\expandafter{\expandafter2\string-4\smallskip} & \textbf{Legitimate} & \textbf{Malicious} & \textbf{Summary}\tabularnewline             
		\noalign{\smallskip} \Xhline{2\arrayrulewidth} \noalign{\smallskip}
		
		\textbf{Apache} & 2911 & 37 & 2948 \tabularnewline
		
		\textbf{Postfix} & 179 & 7 &  186 \tabularnewline
		
		\textbf{Other Traffic} & 2637 &  -- & 2637 \tabularnewline
		
		\midrule
		
		\textbf{Summary} & 5727 & 44 & 5771 \tabularnewline
		\bottomrule
	\end{tabular}
	\caption{ASNM-CDX-2009 dataset distribution.} \label{tab:cdx-dataset-distribution}
	
\end{table}
A distribution of malicious and legitimate samples across obtained dataset is presented in \autoref{tab:cdx-dataset-distribution}.
Beside two services that contained buffer overflow vulnerabilities, our dataset also contains samples representing other network traffic, which we consider as legitimate since no match of its metadata with SNORT log was determined.

\paragraph{Labeling}
ASNM-CDX-2009 dataset contains two types of labels that are enumerated by increasing order of their granularity in the following:
\begin{itemize}
	\item \textbf{label\_2}: is a two-class label, which indicates whether an actual sample represents a network buffer overflow attack or legitimate traffic.
	
	\item \textbf{label\_poly}: is composed of two parts that are delimited by a separator: 
	(a) a two-class label where legitimate and malicious communications are represented by symbols 0 and 1, respectively, and 
	(b) an acronym of network service. 
	This label represents the type of communication on a particular network service.        
\end{itemize} 
This dataset was for the first time used and evaluated in~\cite{homoliakasnm2013}.

\subsection{ASNM-TUN Dataset}\label{sec:desc-TUN}
\begin{figure}
	\centering
	\includegraphics[width=0.98\columnwidth]{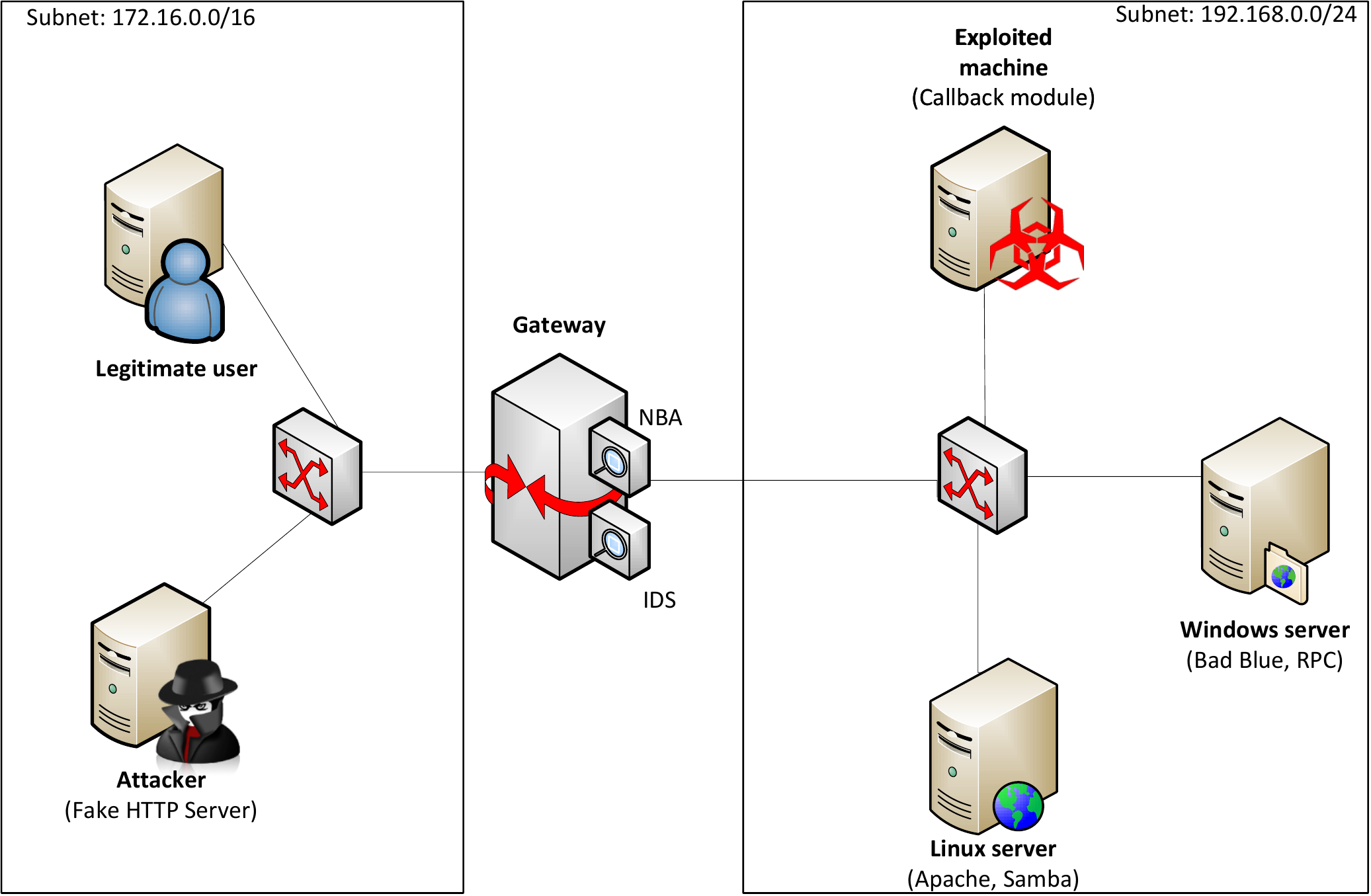}
	\caption{A setup of virtual network used in ASNM-TUN dataset.}\label{fig:networkDraft}
\end{figure}

ASNM-TUN\footnote{The name is derived from \textbf{TUN}elling obfuscations.} dataset was build in laboratory conditions\footnote{Note that part of the legitimate connections was extracted from anonymized metadata collected from a real network.} using a custom virtual network architecture (see \autoref{fig:networkDraft}), where we simulated malicious TCP connections on a few selected vulnerable network services.
\begin{table}
	\begin{center}        
		
		\begin{tabular}{>{\raggedleft}m{2.2cm}>{\centering}m{2.5cm}>{\centering}p{1.0cm}}
			\toprule
			\textbf{Service} & \textbf{CVE} & \textbf{CVSS} \tabularnewline
			\midrule
			
			\textbf{Apache Tomcat   } & 2002-0082  &  7.5 \tabularnewline
			\textbf{BadBlue         } & 2007-6377 &  7.5 \tabularnewline               
			\textbf{DCOM RPC        } & 2003-0352 &  7.5 \tabularnewline                    
			\textbf{Samba           } & 2003-0201 &  10.0 \tabularnewline                
			
			\bottomrule  
		\end{tabular}
		\caption{A list of vulnerable services in ASNM-TUN dataset.}\label{tab:tun-dataset-services}
	\end{center}        
\end{table} 
The selected vulnerabilities are presented in \autoref{tab:tun-dataset-services}, which also contains Common Vulnerabilities and Exposures (CVE) IDs and Common Vulnerability Scoring System (CVSS) values. 
A selection of the vulnerable services was aimed at a high severity of their successful exploitation, namely a presence of buffer overflow vulnerabilities that led to a remote shell code execution through an established backdoor communication, while as a consequence of successful exploitation, the attacker was able to get the root access. 
The details about each vulnerability and its exploitation are briefly described in the following listing:
\begin{compactitem}
	
    \item \textbf{Apache web server with \texttt{mod\_ssl} plugin 2.8.6:}
This attack exploits a buffer overflow vulnerability in \texttt{mod\_ssl} plugin of the Apache web server.
The plugin does not correctly initialize memory in the \texttt{i2d\_SSL\_SESSION} function, which allows a remote attacker to exploit a~buffer overflow vulnerability in order to execute arbitrary code via a~large client certificate that is signed by a trusted Certification Authority, which produces a~large serialized session~\cite{apacheCVE}.
This allows remote code execution and modification of any file on a compromised system~\cite{rapid-7-modssl}. 
The vulnerable versions of the plugin are in range 2.7.1-2.8.6.

\item \textbf{BadBlue web server 2.72b:} 
The second attack exploits a~stack-based buffer overflow vulnerability in \texttt{PassThru} functionality of \texttt{ext.dll} in BadBlue 2.72b and earlier~\cite{rapid-7-badblue}. 
In the attack performing phase, the specially crafted packet with a~long header is sent, which leads to an overflow of processing buffer~\cite{badBlueCVE} 

\item \textbf{Microsoft DCOM RPC:} 
The third attack exploits a vulnerability in Microsoft Windows DCOM Remote Procedure Call (DCOM RPC) service of Microsoft Windows NT 4.0, 2000 (up to Service Pack 4), Server 2003, and XP~\cite{dcomCVE}.
This vulnerability allows a~remote attacker to execute an arbitrary code after a buffer overflow in the DCOM interface.
The vulnerability was originally found by the Last Stage of Delirium research group and has been widely exploited since then~\cite{rapid-7-dcom}.
The vulnerability is well documented, and it was used, for example, by Blaster worm.

\item \textbf{Samba service 2.2.7:} 
The last attack exploits a buffer overflow vulnerability in \texttt{call\_ trans2open} function in \texttt{trans2.c} of Samba 2.2.x before 2.2.8a, 2.0.10, earlier versions than 2.0.x and Samba-TNG before 0.3.2~\cite{sambaCVE}.        
This vulnerability allows a~remote attacker to execute arbitrary code.
An exploit code sends malformed packets to a~remote server in batches~\cite{rapid-7-samba}.
Packets differ in one shell-code address only because the return address depends on versions of Samba and host operating systems.    
\end{compactitem}

\paragraph{Adversarial Modifications}
We employed tunneling of malicious network traffic inside of HTTP and HTTPS protocols, serving as obfuscation techniques when exploiting vulnerable services. 
The tunneling obfuscation modifies $P_{c}$ and $P_{s}$ packet sets (see \autoref{sec:tcp-con}) of the original malicious connection $c_{m}$ by wrapping each original packet into a new one.
Assuming the background from \autoref{sec:adversarial-obfuscations}, the tunneling (i.e., wrapping) may cause fragmentation of IP packets, and thus it can also modify the number of packets in both packet sets $P_{c}$ and $P_{s}$.
Also, the obfuscation modifies IP addresses ($ip_c, ip_s$)  and ports ($p_c, p_s$) of the original connection.
Symbols of the packet tuple whose values are sensitive to the obfuscation
include all defined fields, as tunneling obfuscation creates new TCP/IP stack with unique values of L2, L3, L4 headers as well as new content of application layer data. 
All these modifications, especially modifications of $P_{c}$ and $P_{s}$ of the connection $c_{m}$, cause alteration of the original network connection features' values (see \autoref{sec:adversarial-obfuscations}).

For the purpose of simulating real network conditions, we executed each malicious and legitimate network communication four times with four different network traffic modifications.
Network traffic modifications differ in the alteration degree of the network traffic, and we divide them into four categories:
\begin{enumerate}
\item[(a)] \textbf{No Modification:} The first category represents reference output without any modification. 
All experiments ran on the same host machine to minimize deviations among different tests.

\item[(b)] \textbf{Traffic Shaping:} The second category is dedicated to simulation of traffic shaping. Therefore, all packets were forwarded with higher time delays. For this purpose, the special gateway machine with a limited processor's performance was used. This machine was also fully loaded to emulate slower packets processing than in the first scenario.

\item[(c)] \textbf{Traffic Policing:} The third category is supposed to simulate traffic policing when some of the packets were dropped during the processing on the network gateway node. In this case, a custom packet dropper was used on the gateway node, and 25\% of packets were dropped, resulting in output which contained re-transmitted packets.

\item[(d)] \textbf{Corrupted Traffic:} The fourth category represents transmission on an unreliable network channel; thus, 25\% of packets were corrupted during processing on the network gateway node.    
\end{enumerate}

\paragraph{Legitimate Network Traffic}
Legitimate samples of the dataset were collected from two sources. 
The first source represents a legitimate traffic simulation in our virtual network architecture and also employed network traffic modifications for the purpose of simulating real network conditions. 
As the second source, common usage of selected services was captured in the campus network in accordance with policies in force.
In the obtained data, no content of packets was captured, and all collected metadata was anonymized.
Further, we filter out data matched on high severity alerts by signature-based Network Intrusion Detection Systems (NIDS) Suricata~\cite{suricataWeb} and SNORT~\cite{snortWeb} through Virus Total API.
This step assured that legitimate traffic does not contain any malicious data. 
Note that SNORT was equipped with \textit{Sourcefire VRT} ruleset, and SURICATA utilized \textit{Emerging Threats ETPro} ruleset. 
The final composition of the dataset after extraction of ASNM features is depicted in \autoref{tab:TUN-dataset_disribution}. 

\setlength{\tabcolsep}{4pt}
\begin{table}[t]
\centering    

\begin{tabular}{>{\raggedleft}p{2.2cm}c>{\centering}p{1.3cm}>{\centering}p{1.3cm}>{\centering}p{1.3cm}}
	\toprule
	\multirow{2}{3.6cm}{\textbf{~Network Service}} & \multicolumn{4}{c}{\textbf{Count of TCP Connections}} \smallskip \tabularnewline
	\expandafter\cline\expandafter{\expandafter2\string-5\smallskip} & \textbf{Legitimate} & \textbf{Direct Attacks} & \textbf{Obfuscated Attacks} & \textbf{Summary}\tabularnewline             
	\noalign{\smallskip} \Xhline{2\arrayrulewidth} \noalign{\smallskip}
	
	\textbf{Apache} & 38 & 102 & 61 & 201\tabularnewline
	
	\textbf{BadBlue} & 95 & 4 & 10 & 109\tabularnewline
	
	\textbf{DCOM RPC} & 4 & 4 & 8 & 16\tabularnewline
	
	\textbf{Samba} & 15 & 20 & 8 & 43\tabularnewline
	
	\textbf{Other Traffic} & 25 & -- & -- & 25\tabularnewline
	
	\midrule
	
	\textbf{Summary} & 177 & 130 & 87 & 394\tabularnewline
	\bottomrule
\end{tabular}

\caption{ASNM-TUN dataset distribution.}\label{tab:TUN-dataset_disribution}

\end{table} \hspace{-0.5cm} \setlength{\tabcolsep}{1.4pt}

\paragraph{Labeling}
ASNM-TUN dataset  contains four types of labels that are enumerated by increasing order of their granularity in the following:
\begin{itemize}

\item \textbf{label\_2:} is a two-class label, which indicates whether an actual sample represents a network buffer overflow attack or a legitimate communication.

\item \textbf{label\_3:}  is a three-class label, which distinguishes among legitimate traffic (symbol 3), direct attacks (symbols 1), and obfuscated network attacks (symbol 2).

\item \textbf{label\_poly:} is a label that is composed of 2 parts: (a) a three-class label, and 
(b) acronym of a network service. 
This label represents a type of communication on a particular network service.

\item \textbf{label\_poly\_s:} is composed of 3 parts: 
(a) a three-class label, 
(b) an acronym of network service, and 
(c) a network modification technique employed. 
This label has almost the same interpretation as the previous one, but in addition, it introduces a network modification technique employed (identified by a letter from the previous listing).
\end{itemize}

\paragraph{Testing with Signature-Based NIDS}
To investigate the effect of the tunneling obfuscation on signature-based NIDSs, we performed detection by SNORT and SURICATA through VirusTotal API~\cite{virusTotalUrl}.
SNORT was equipped with Sourcefire VRT ruleset, and SURICATA utilized Emerging Threats ETPro ruleset. 
The results of direct attacks' detection by both NIDSs are shown in \autoref{tunNidsSNORTSuricataDirect}.
Note that high priority rules detected 93 direct attacks on Apache service in both NIDSs, but 4 undetected direct attacks occurred almost at the same time as some of the detected attack instances, and hence, we consider them as a part of other detected direct attacks.
Also, we can see that five instances of direct attacks were not detected by SNORT nor SURICATA.
These five instances utilized network traffic modifications (c) and (d), which likely influenced the detection rate of both NIDSs; hence, they give an intuition for the adversarial obfuscation techniques utilized in the last ASNM dataset (see \autoref{sec:desc-NPBO}). 
The resulting detection rates of direct attacks look the same in both NIDSs, but there were differences in fired alerts during the exploitation of Apache service.
Unlike SNORT, SURICATA had not detected any occurrence of buffer overflow, nor shellcode,
nor remote command execution but instead fired high priority alerts related to potential corporate privacy violation: 
\begin{itemize}
\item
\begingroup
\fontsize{8pt}{10pt}
\begin{verbatim}
ET POLICY 
Possible SSLv2 Negotiation in Progress 
Client Master Key SSL2_RC4_128_WITH_MD5,                    
\end{verbatim}
\endgroup
\end{itemize}
which we decided to consider as correct detection.
If we would not consider them as correctly detected, then SURICATA was not detecting any direct attack on the Apache service. 

\setlength{\tabcolsep}{4pt}
\begin{table}[t]
	\centering       
	
	\begin{centering}        
		\subfloat[SNORT]{
			\protect\centering{}
			
			\begin{tabular}{>{\raggedleft}p{2.6cm}c>{\centering}p{1.2cm}>{\centering}p{1.5cm}}
				
				\toprule
				
				\multirow{2}{3.6cm}{\textbf{~}} & \multicolumn{3}{c}{\textbf{Direct Attacks}}		
				\tabularnewline
				
				\expandafter\cline\expandafter{\expandafter2\string-4\smallskip} 
				& \textbf{Detected} & \textbf{Total} & \textbf{\%} \tabularnewline             
				
				\noalign{\smallskip} \Xhline{2\arrayrulewidth} \noalign{\smallskip}
				
				\textbf{Apache} & ~~~~93 $+$4 & 102 & 95.10\% \tabularnewline
				
				\textbf{BadBlue} & 4 & 4 & 100.00\% \tabularnewline
				
				\textbf{DCOM RPC} & 4 & 4 & 100.00\% \tabularnewline
				
				\textbf{Samba} & 20 & 20 & 100.00\% \tabularnewline
				
				\Xhline{1\arrayrulewidth}\noalign{\smallskip}
				
				\textbf{Overall Detection} & 125 & 130 & 96.15\% \tabularnewline
				\textbf{ADR$^*$ per Service} &  &  & 98.77\% \tabularnewline
				\bottomrule
				\multicolumn{4}{r}{$^*$Average detection rate.}
			\end{tabular}     
			
		}
	\end{centering}
	
	\begin{centering}		
		\subfloat[SURICATA]{
			\centering
			
			\begin{tabular}{>{\raggedleft}p{2.6cm}c>{\centering}p{1.2cm}>{\centering}p{1.5cm}}
				
				\Xhline{2\arrayrulewidth} \noalign{\smallskip}
				\multirow{2}{3.6cm}{\textbf{~}} & \multicolumn{3}{c}{\textbf{Direct Attacks}} 
				\tabularnewline
				
				\expandafter\cline\expandafter{\expandafter2\string-4\smallskip} 
				& \textbf{Detected} & \textbf{Total} & \textbf{\%} \tabularnewline             
				
				\Xhline{2\arrayrulewidth} \noalign{\smallskip}
				
				\textbf{Apache} & ~~~~93 $+$4 & 102 & 95.10\% \tabularnewline
				
				\textbf{BadBlue} & 4 & 4 & 100.00\% \tabularnewline
				
				\textbf{DCOM RPC} & 4 & 4 & 100.00\% \tabularnewline
				
				\textbf{Samba} & 20 & 20 & 100.00\% \tabularnewline
				
				\Xhline{1\arrayrulewidth}\noalign{\smallskip}
				
				\textbf{Overall Detection} & 125 & 130 & 96.15\% \tabularnewline
				\textbf{ADR$^*$ per Service} &  &  & 98.77\% \tabularnewline
				\Xhline{2\arrayrulewidth} 
				\multicolumn{4}{r}{$^*$Average detection rate.}
			\end{tabular}     
			
		}
	\end{centering}  
	\caption{Detection of direct attacks in ASNM-TUN dataset by SNORT and SURICATA.}
	\label{tunNidsSNORTSuricataDirect}
	
\end{table} \setlength{\tabcolsep}{1.4pt} 

\medskip \noindent
Next, we have performed exploitation of each vulnerable service using the tunneling obfuscation, while scanning the network by aforementioned NIDSs.
The obtained results are depicted in \autoref{tun_NIDS_Snort_Suricata_obfus}, which distinguishes between tunneling obfuscation performed through HTTP and HTTPS protocols.
\setlength{\tabcolsep}{2pt}
\begin{table*}[h]
	\centering{}       
	\smallskip\smallskip
	
	\begin{centering}        
		\subfloat[SNORT]{
			\protect\centering{}
			
			\begin{tabular}{>{\raggedleft}p{1.7cm}ccc |c ccc |c ccc}
				
				\Xhline{2\arrayrulewidth} \noalign{\smallskip}
				
				\multirow{3}{1.7cm}{\textbf{~Service}} & \multicolumn{11}{c}{\textbf{Obfuscated Attacks}}		
				\tabularnewline
				
				\expandafter\cline\expandafter{\expandafter2\string-12\smallskip} 
				
				~ & \multicolumn{3}{c}{\textbf{HTTP}} & ~ & \multicolumn{3}{c}{\textbf{HTTPS}} & ~ & \multicolumn{3}{c}{\textbf{All}} 
				\tabularnewline
				
				~ & \textbf{Detected} & \textbf{Total} & \textbf{\%} & & \textbf{Detected} & \textbf{Total} & \textbf{\%} & & \textbf{Detected} & \textbf{Total} & \textbf{\%}
				\tabularnewline             
				
				\Xhline{2\arrayrulewidth} \noalign{\smallskip}
				
				\textbf{Apache} & 0  & 4 & 0.00 &   & ~~~~~51 $+$6  & 57 & 100.00& &  57  & 61 & 93.40\tabularnewline
				
				\textbf{BadBlue} & 3 & 6 & 50.00 &  & 2  & 4  & 50.00 & &  5  & 10 & 50.00 \tabularnewline
				
				\textbf{DCOM} & 0 & 4 &     0.00 &  & 3  & 4  & 75.00 & &  3  & 8 & 37.50 \tabularnewline
				
				\textbf{Samba} & 0 & 4 &    0.00 &  & 2  & 4  & 50.00 & &  2  & 8 & 25.00 \tabularnewline
				
				\Xhline{1\arrayrulewidth} \noalign{\smallskip}
				
				\textbf{Summary} & 3 & 18 & 16.67 & &  64  & 69 & 92.75 &  & 67 & 87 & 77.01  \tabularnewline
				\textbf{ADR$^*$} &  &  & 12.50 &   &  &  & 68.75 &  &  &  & 51.49 \tabularnewline
				\Xhline{2\arrayrulewidth} 
				\multicolumn{4}{r}{$^*$Average detection rate per class.}
			\end{tabular}      
			
		}
	\end{centering}
	
	\begin{centering}
		\smallskip
		\subfloat[SURICATA]{
			\protect\centering{}
			
			\begin{tabular}{>{\raggedleft}p{1.7cm}ccc |c ccc |c ccc}
				
				\Xhline{2\arrayrulewidth} \noalign{\smallskip}
				
				\multirow{3}{1.7cm}{\textbf{~Service}} & \multicolumn{11}{c}{\textbf{Obfuscated Attacks}}		
				\tabularnewline
				
				\expandafter\cline\expandafter{\expandafter2\string-12\smallskip} 
				
				~ & \multicolumn{3}{c}{\textbf{HTTP}} & ~ & \multicolumn{3}{c}{\textbf{HTTPS}} & ~ & \multicolumn{3}{c}{\textbf{All}} 
				\tabularnewline
				
				~ & \textbf{Detected} & \textbf{Total} & \textbf{\%} & & \textbf{Detected} & \textbf{Total} & \textbf{\%} & & \textbf{Detected} & \textbf{Total} & \textbf{\%}
				\tabularnewline             
				
				\Xhline{2\arrayrulewidth} \noalign{\smallskip}
				
				\textbf{Apache} & 0  & 4 & 0.00 &   & ~~~~~50 $+$3  & 57 & 92.98& &  53  & 61 & 86.89\tabularnewline
				
				\textbf{BadBlue} & 3 & 6 & 50.00 &  & 0  & 4  & 0.00 & &  3  & 10 & 30.00 \tabularnewline
				
				\textbf{DCOM} & 0 & 4 &     0.00 &  & 0  & 4  & 0.00 & &  0  & 8 & 0.00 \tabularnewline
				
				\textbf{Samba} & 0 & 4 &    0.00 &  & 0  & 4  & 0.00 & &  0  & 8 & 0.00 \tabularnewline
				
				\Xhline{1\arrayrulewidth} \noalign{\smallskip}
				
				\textbf{Summary} & 3 & 18 & 16.67 & &  53  & 69 & 76.81 &  & 56 & 87 & 64.37  \tabularnewline
				\textbf{ADR$^*$} &  &  & 12.50 &   &  &  & 23.25 &  &  &  & 29.22 \tabularnewline
				\Xhline{2\arrayrulewidth} 
				\multicolumn{4}{r}{$^*$Average detection rate per class.}
			\end{tabular} 
			
		}
	\end{centering}  
	\caption{Detection of obfuscated attacks in ASNM-TUN dataset by SNORT and SURICATA.}
	\label{tun_NIDS_Snort_Suricata_obfus}
	
\end{table*} \hspace{-0.5cm} \setlength{\tabcolsep}{1.4pt}

We can see that an average detection rate per service is significantly lower for obfuscated attacks than in the case of direct attacks, and thus tunneling obfuscation was partially capable of evading detection by utilized NIDSs.
Regarding tunneling through the HTTP protocol, both SNORT and SURICATA achieved the same low detection rate for all classes of attacks.

The situation is slightly different for the case of tunneling through the HTTPS protocol.
The SNORT achieved an average detection rate (ADR) per class equal to $68.75\%$ and SURICATA only $23.25\%$.
We found out the same fact about high priority rules fired by SURICATA on exploitation of Apache service as in the case of direct attacks detection -- neither buffer overflow, nor shellcode, nor remote command execution rules were matched, and thus we decided to accept the previously mentioned potential corporate privacy violation alert as correct detection again. 
If we would not accept it, then SURICATA were not detected any tunneled attack on Apache service.
Also note that SURICATA fired one non-high-priority alert classified as potentially bad traffic in several instances of attacks tunneled through HTTPS, which exploited BadBlue, DCOM and Samba services:
\begin{itemize}
	\item
	\begingroup
	\fontsize{8pt}{10pt}
	\begin{verbatim}
	ET POLICY 
	FREAK Weak Export Suite 
	From Client (CVE-2015-0204).
	\end{verbatim}
	\endgroup
\end{itemize}
But we have not considered it as correct detection due to the low priority of the alert as well as the scope of corresponding CVE-2015-0204 is only related to the client code of OpenSSL.
The plus notation in \autoref{tun_NIDS_Snort_Suricata_obfus}, alike in  \autoref{tunNidsSNORTSuricataDirect}, denotes undetected attacks that occurred almost at the same time as some other correctly detected attacks, and thus are considered as their parts. 
Concluding the results of NIDSs detection, we can state that the proposed tunneling obfuscation technique was successful in evading the NIDSs used since a high number of obfuscated attacks were not detected in comparison to the case where obfuscations were not employed.
On the other hand, we emphasize that SNORT has detected the most of direct attacks on Apache service even though it was encrypted.
This indicates that VirusTotal may utilize a very paranoid rule set, which causes false positives.
Hence, the results of the analysis through VirusTotal API are arguable.

\subsection{ASNM-NPBO Dataset}\label{sec:desc-NPBO}
\begin{figure}
	\centering
	\includegraphics[width=0.95\columnwidth]{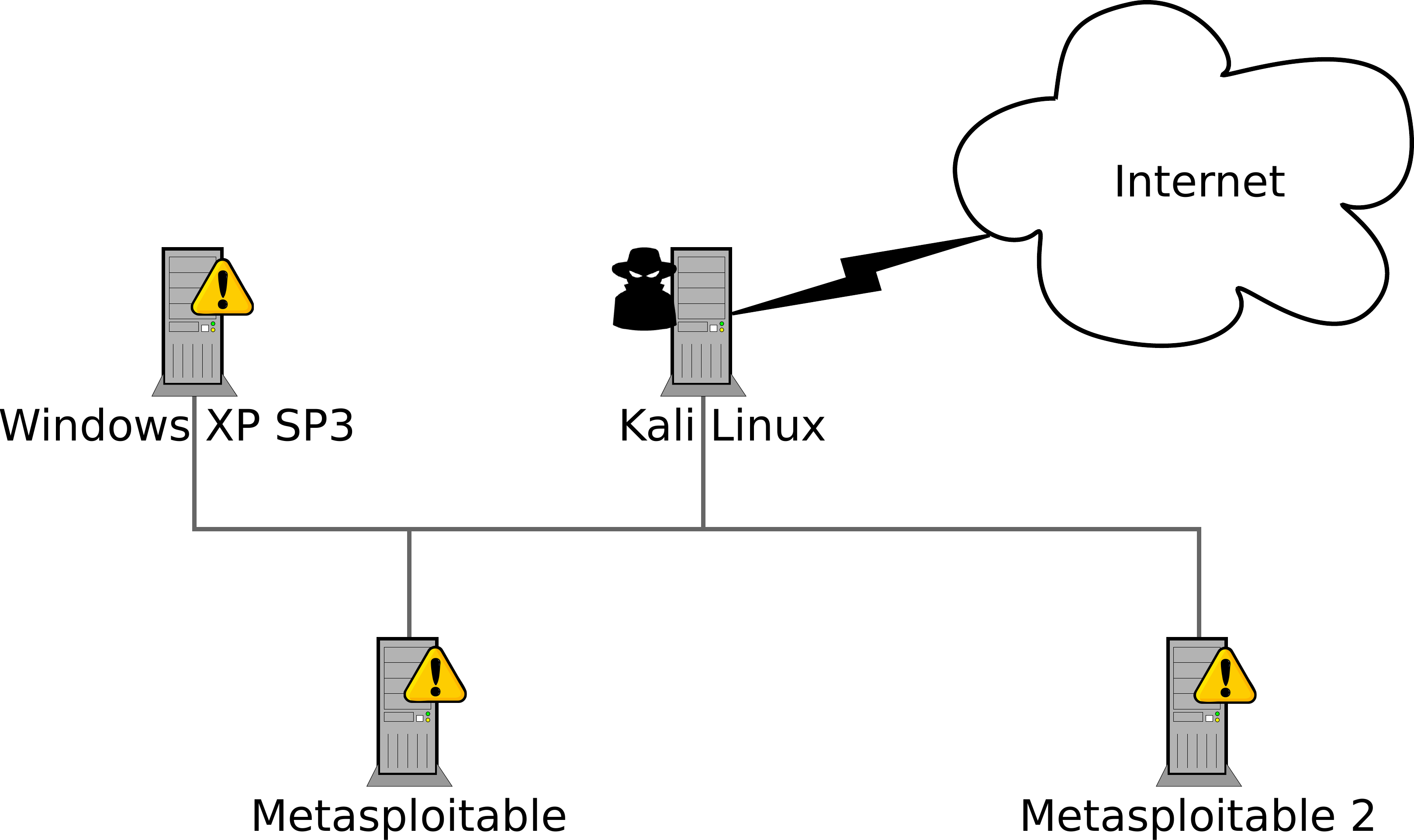}
	\caption{A setup of virtual network used in ASNM-NPBO dataset.}\label{fig:network-NPBO}
\end{figure}

ASNM-NPBO\footnote{The name is derived from \textbf{N}on-\textbf{P}ayload-\textbf{B}ased \textbf{O}bfuscations.} dataset was built in laboratory conditions using a virtual network architecture (see \autoref{fig:network-NPBO}) consisting of three vulnerable machines and the attacker's machine.
\begin{table}[b]
	\begin{center}     
		
		\begin{tabular}{>{\raggedleft}m{3.3cm}>{\centering}m{2.5cm}>{\centering}p{1.0cm}}
			\Xhline{2\arrayrulewidth} \noalign{\smallskip}
			\textbf{Service} & \textbf{CVE} & \textbf{CVSS} \tabularnewline
			\noalign{\smallskip} \Xhline{1\arrayrulewidth} \noalign{\smallskip}

			\textbf{Apache Tomcat   } & 2009-3843 &  10.0 \tabularnewline
			\textbf{DistCC service         } & 2004-2687  &  9.3 \tabularnewline               
			\textbf{MSSQL           } & 2000-1209 &  10.0 \tabularnewline                    
			\textbf{PostgreSQL      } & 2007-3280 &  9.0 \tabularnewline                
			\textbf{Samba service          } & 2007-2447 &  6.0 \tabularnewline                
			\textbf{Server service of Windows         } & 2008-4250 &  10.0 \tabularnewline                
			
			\Xhline{2\arrayrulewidth}  
		\end{tabular}
		\caption{A list of vulnerable services in ASNM-NPBO dataset.}\label{tab:NPBO-services}
		
	\end{center}        
\end{table}
\setlength{\tabcolsep}{1.4pt}     
All virtual machines were configured with private static IP addresses in order to enable easy automation of the whole exploitation process. 
Our testing network infrastructure consisted of the attacker's machine equipped with Kali Linux and vulnerable machines that were running Metasploitable 1, 2~\cite{metasploitable}, and
Windows XP with SP~3. 
We aimed at the selection of vulnerable services with the high severity of their successful exploitation leading to remote shell code execution through an established backdoor communication.
All selected vulnerable services are depicted in \autoref{tab:NPBO-services}, which also contains CVE IDs and CVSS severity score values.
The details about each vulnerability and its exploitation are briefly described in the following:
\begin{itemize}
	\item \textbf{Apache Tomcat 5.5:} 
	First, a dictionary attack was executed in order to obtain access credentials into the application manager instance~\cite{ms-tomcat_mgr_login}.
	Further, the server's application manager was exploited for the transmission and execution of malicious code~\cite{ms-tomcat_mgr_deploy}.        
	
	\item \textbf{Microsoft SQL Server 2005:}       
	A dictionary attack was employed to obtain access credentials of MSSQL user~\cite{ms-mssql_login} and then the procedure \texttt{xp\_cmd\-shell} enabling the execution of an arbitrary code was exploited~\cite{ms-mssql_payload}.         
	
	\item \textbf{Samba 3.0.20-Debian:}
	A vulnerability in Samba service enabled the attacker of arbitrary command execution, which exploited MS-RPC functionality when \texttt{username\_map\_script}~\cite{ms-usermap_script} was allowed in the configuration.
	There was no need for authentication in this attack.
	
	\item \textbf{Server Service of Windows XP:}
	The server service enabled the attacker of arbitrary code execution through crafted RPC request resulting in stack overflow during path canonicalization~\cite{ms-ms08_067_netapi}.              
	
	\item \textbf{PostgreSQL 8.3.8:}
	A dictionary attack was executed in order to obtain access credentials  into the PostgreSQL instance~\cite{ms-postgres_login}.
	Standard PostgreSQL Linux installation had write access to \texttt{/tmp} directory, and it could call user-defined functions (UDF).
	UDFs utilized shared libraries located on an arbitrary path (e.g., \texttt{/tmp}). 
	An attacker exploited this fact and copied its own UDF code to \texttt{/tmp} directory and then executed it~\cite{ms-postgres_payload}.
	
	\item \textbf{DistCC 2.18.3:}
	A vulnerability enabled the attacker remote execution of an arbitrary command through compilation jobs that were executed on the server without any permission check~\cite{ms-distcc_exec}. 
	
\end{itemize}

\paragraph{Adversarial Modifications}
We proposed several non-payload-based obfuscation techniques~\cite{homoliak2016exploitation} when exploiting vulnerable network services as well as during the execution of legitimate communications on the services. 
The proposed non-payload-based obfuscation techniques are described in \autoref{tab:NPBO-obfuscation-techniques}, 
Assuming the background from \autoref{sec:adversarial-obfuscations}, the proposed non-payload-based obfuscation techniques can modify $P_{c}$ and $P_{s}$ packet sets of the original connection
$c_{m}$ by insertion, removal and transformation of the packets.
Symbols of the~packet tuple (see \autoref{tab:Packet_tuple_items}) whose values are sensitive to the obfuscations
include: $t,size,$ $ip_{off},$ $ip_{sum},$ $tcp_{sum},$ $tcp_{seq},$
$tcp_{ack},$ $tcp_{off},$ $tcp_{flags},$ $tcp_{win},$ $tcp_{urp}$
and $data$.\footnote{Note the $data$ field is sensitive to the obfuscations only in the manner of damaging or splitting the original packet's data.} 
The modifications of $P_{c}$ and $P_{s}$ of the connection $c_{m}$
can cause alteration of the original network connection features' values $F^{m}$ to new ones (see \autoref{sec:adversarial-obfuscations}). 

\setlength{\tabcolsep}{4pt}    
\begin{table}[t]
	\footnotesize{
		\begin{center}
			
			\begin{tabular}{p{2.5cm} p{5.0cm} m{0.3cm}}
				\toprule
				\textbf{Technique} & \textbf{Parametrized Instance} & \textbf{ID}\\            
				\noalign{\smallskip} \Xhline{2\arrayrulewidth} \noalign{\smallskip}
				
				\multirow{3}{4cm}[-0.2cm]{{\textbf{Spread out packets\\in time} }} & {$\bullet$ constant delay: 1s} & {(a)} \\
				
				& {$\bullet$ constant delay: 8s } & {(b)}\\
				
				& {$\bullet$ normal distribution of delay with 5s mean 2.5s standard deviation (25\% correlation)} & {(c)}\\
				\midrule 
				
				\textbf{Packets' loss} & {$\bullet$ 25\% of packets} & {(d)}\\
				\midrule 
				\multirow{3}{4cm}{{\textbf{Unreliable network\\channel simulation}}} & {$\bullet$ 25\% of packets damaged} & {(e)}\\
				
				& {$\bullet$ 35\% of packets damaged} & {(f)}\\
				
				& {$\bullet$ 35\% of packets damaged with~25\% correlation} & {(g)}\\
				\midrule 
				\textbf{Packets' duplication} & {$\bullet$ 5\% of packets} & {(h)}\\
				\midrule 
				\multirow{2}{4cm}[-0.2cm]{\textbf{Packets' order\\modifications}} & {$\bullet$ reordering of 25\% packets;  reordered packets are sent with~10ms delay and 50\%~correlation} & {(i)}\\
				
				& {$\bullet$ reordering of 50\% packets;  reordered packets are sent with~10ms delay and 50\%~correlation} & {(j)}\\
				\midrule 
				\multirow{4}{4cm}{\textbf{Fragmentation}} & {$\bullet$ MTU 1000} & {(k)}\\
				
				& {$\bullet$ MTU 750} & {(l)}\\
				
				& {$\bullet$ MTU 500} & {(m)}\\
				
				& {$\bullet$ MTU 250} & {(n)}\\
				\midrule 
				\multirow{3}{4cm}[-0.5cm]{\textbf{Combinations}} & {$\bullet$ normal distribution delay ($\mu = 10ms$, $\sigma = 20ms$) and 25\% correlation; loss:
					23\% of packets; corrupt: 23\% of packets; reorder:~23\% of packets} & {(o)}\\
				
				& {$\bullet$ normal distribution delay ($\mu = 7750ms$, $\sigma = 150ms$) and 25\% correlation; loss:~0.1\% of packets;
					corrupt:~0.1\% of packets; duplication:~0.1\% of packets; reorder: 0.1\% of packets} & {(p)}\\
				
				& {$\bullet$ normal distribution delay ($\mu = 6800ms$, $\sigma = 150ms$) and 25\% correlation; loss: 1\% of packets;
					corrupt:~1\% of packets; duplication:~1\% of packets; reorder~1\% of packets} & {(q)}\\
				\bottomrule
				
			\end{tabular}        
		\end{center}
	}
	\caption{Non-payload-based obfuscation techniques with parameters and IDs.}
	\label{tab:NPBO-obfuscation-techniques}
	\vspace{-0.6cm}
	
\end{table}
\setlength{\tabcolsep}{1.4pt}

\begin{figure}[t]
	\centering
	\vspace{-0.3cm}
	\includegraphics[width=\columnwidth]{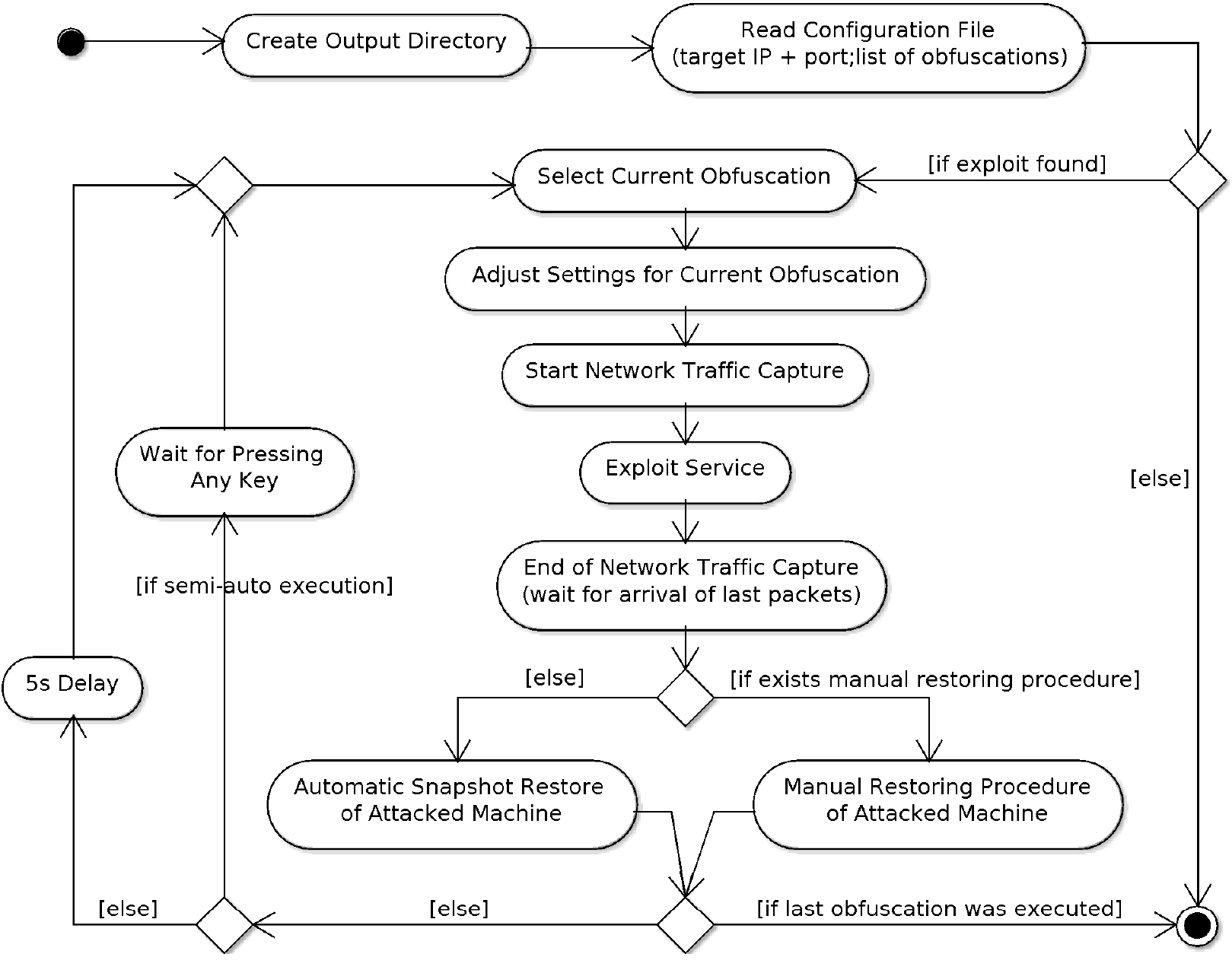}
	\caption{Behavioral state diagram of the obfuscation tool.}
	
	\label{fig:BehavioralStateDiagram}    
\end{figure}

\medskip \noindent
Then we built an obfuscation tool~\cite{ih2019nonpayload} that morphs network characteristics of a TCP connection at network and transport layers of the TCP/IP stack by applying one or a combination of several non-payload-based obfuscation techniques.
Execution of direct communications (non-obfuscated ones) is also supported by the tool as well as capturing network traffic related to communication. 
The tool is capable of automatic/semi-automatic run and restoring of all modified system settings and consequences of attacks/legitimate communications on a target machine.
After the successful execution of each desired obfuscation on the selected service, the output of the tool contains several network packet traces associated with pertaining obfuscations.
The behavioral state diagram of the obfuscation tool is depicted in Figure~\autoref{fig:BehavioralStateDiagram}.

We applied our obfuscation tool for automatic exploitation of all enumerated vulnerable services while using the proposed obfuscations. 
When exploitation leading to a remote shell was successful, simulated attackers performed simple activities involving various shell commands (such as listing directories, opening, and reading files). 
An average number of issued commands was around 10, and text files of up to 50kB were opened/read.
Note that we labeled each TCP connection representing dictionary attacks as legitimate ones due to two reasons: 1) from the behavioral point of view, they independently appeared just as unsuccessful authentication attempts, which may occur in legitimate traffic as well, 2) more importantly, we employed ASNM features whose subset involves context of an analyzed TCP connection for their computation -- i.e., ASNM features capture relations to other TCP connections initiated from/to a corresponding service.

\paragraph{Legitimate Network Traffic}
The legitimate samples of this dataset were collected from two sources:
\begin{itemize}
	\item A common usage of all previously mentioned services was obtained in an anonymized form, excluding the payload, from a real campus network in accordance with policies in force. 
	Analyzing packet headers, we observed that a lot of expected legitimate traffic contained malicious activity, as many students did not care about up-to-date software.  
	Therefore, we filtered out network connections yielding high and medium severity alerts by signature-based NIDS -- Suricata and SNORT -- through Virus Total API~\cite{virusTotalUrl}.
	
	\item The second source represented legitimate traffic simulation in our virtual network architecture and also employed all of our non-payload-based obfuscations for the purpose of partially addressing overstimulation in adversarial attacks against IDS~\cite{corona2013adversarial}, and thus making the classification task more challenging.
	However, only 109 TCP connections were obtained from this stage, which was also caused by the fact that services such as Server and DistCC were hard to emulate.\footnote{Note that additionally to those 109 TCP connections that were explicitly simulated, other 2252 TCP connections from obfuscated dictionary attacks were also considered as legitimate, and thus also helped in achieving a resistance against the overstimulation attacks.}
	Simulation of legitimate traffic was aimed at various \textit{SELECT} and \textit{INSERT} statements when interacting with the database services (i.e., PostgreSQL, MSSQL); several \textit{GET} and \textit{POST} queries to our custom pages as well as downloading of high volume data when interacting with our HTTP server (i.e., Apache Tomcat); and several queries for downloading and uploading small files into Samba share. 
\end{itemize}
The class distribution of the final dataset after extraction of ASNM features is summarized in \autoref{tab:NPBO-Dataset}
\setlength{\tabcolsep}{4pt}
\begin{table}[b]
	\begin{center}        
		\footnotesize{
			\begin{tabular}{>{\raggedleft}p{2.1cm}c>{\centering}p{1.3cm}>{\centering}p{1.3cm}>{\centering}p{1.2cm}}
				\Xhline{2\arrayrulewidth} \noalign{\smallskip}
				\multirow{2}{3.6cm}{\textbf{~Network Service}} & \multicolumn{4}{c}{\textbf{Count of TCP Connections}} \smallskip \tabularnewline
				\expandafter\cline\expandafter{\expandafter2\string-5\smallskip} & \textbf{Legitimate} & \textbf{Direct Attacks} & \textbf{Obfuscated Attacks} & \textbf{Summary}\tabularnewline             
				\noalign{\smallskip} \Xhline{2\arrayrulewidth} \noalign{\smallskip}
				
				\textbf{Apache Tomcat} & 809 & 61 & 163 & 1033\tabularnewline
				
				\textbf{DistCC} & 100 & 12 & 23 & 135\tabularnewline
				
				\textbf{MSSQL} & 532 & 31 & 103 & 666\tabularnewline
				
				\textbf{PostgreSQL} & 737 & 13 & 45 & 795\tabularnewline
				
				\textbf{Samba} & 4641 & 19 & 44 & 4704\tabularnewline             
				
				\textbf{Server} & 3339 & 26 & 100 & 3465\tabularnewline             
				
				\textbf{Other Traffic} & 647 & -- & -- & 647\tabularnewline
				
				\midrule
				\textbf{Summary} & 10805 & 162 & 478 & 11445\tabularnewline
				\Xhline{2\arrayrulewidth} 
			\end{tabular}
		}
	\end{center}        
	\caption{ANSM-NPBO dataset distribution.}
	\label{tab:NPBO-Dataset}
\end{table}
\setlength{\tabcolsep}{1.4pt}

\paragraph{Labeling}
ASNM-NPBO dataset contains four types of labels that are enumerated by increasing order of their granularity in the following:
\begin{itemize}
	\item \textbf{label\_2:} is a two-class label, which indicates whether an actual sample represents a network buffer overflow attack or a legitimate communication.
	
	\item \textbf{label\_3:} is a three-class label, which distinguishes among legitimate traffic (symbol 3), direct attacks (symbols 1), and obfuscated network attacks (symbol 2).
	
	\item \textbf{label\_poly:} is a label that is composed of 2 parts: (a) a three-class label, and 
	(b) acronym of a network service. 
	This label represents a type of communication on a particular network service.    
	
	\item \textbf{label\_poly\_o} is the last label, which  is composed of 3 parts: 
	(a) three-class label, 
	(b) employed obfuscation technique, and 
	(c) acronym of network service. 
	The label has almost the same interpretation as label\_poly but moreover introduces obfuscation technique employed (identified by ID from \autoref{tab:NPBO-obfuscation-techniques}) into all obfuscated attack samples.        
\end{itemize}

\paragraph{Testing with Signature-Based NIDS}
To investigate the effect of the proposed non-payload-based obfuscations on signature-based NIDSs, we performed detection by SNORT and SURICATA in a similar manner as we did in the case of the tunneling obfuscations (see \autoref{sec:desc-TUN}), while the same ruleset was employed.

First, we let NIDSs inspect direct attacks that exploit the current network vulnerabilities.
The results of the inspection summarize the detection properties of SNORT and SURICATA, and are depicted in \autoref{non-payload-NIDS-Snort-Suricata-direct}.
\setlength{\tabcolsep}{4pt}
\begin{table}
	\centering{}       
	\smallskip\smallskip
	
	\begin{centering}        
		\subfloat[SNORT\label{non-payload-direct-SNORT}]{
			\protect\centering{}
			
			\begin{tabular}{>{\raggedleft}p{3.6cm}c>{\centering}p{1.2cm} r}
				
				\Xhline{2\arrayrulewidth} \noalign{\smallskip}
				
				\multirow{2}{3.6cm}{\textbf{~}} & \multicolumn{3}{c}{\textbf{Direct Attacks}}        
				\tabularnewline
				
				\expandafter\cline\expandafter{\expandafter2\string-4\smallskip} 
				& \textbf{Detected} & \textbf{Total} & \textbf{\%~~} \tabularnewline             
				
				\Xhline{2\arrayrulewidth} \noalign{\smallskip}
				
				\textbf{Apache Tomcat} & ~~~~~~33 $+$28 & 61 & 100.00 \tabularnewline
				
				\textbf{DistCC} & 12 & 12 & 100.00 \tabularnewline
				
				\textbf{MSSQL} & 31 & 31 & 100.00 \tabularnewline
				
				\textbf{PostgreSQL} & 13 & 13 & 100.00 \tabularnewline             
				
				\textbf{Samba} & 19 & 19 & 100.00 \tabularnewline
				
				\textbf{Server} & 26 & 26 & 100.00 \tabularnewline                                       
				
				\Xhline{1\arrayrulewidth}\noalign{\smallskip}
				
				\textbf{Overall Detection} & 162 & 162 & 100.00 \tabularnewline
				\textbf{ADR$^*$ per Service} &  &  & 100.00 \tabularnewline
				\Xhline{2\arrayrulewidth} 
				\multicolumn{4}{r}{$^*$Average detection rate.}
			\end{tabular}     
			
		}
	\end{centering}
	
	\begin{centering}
		\smallskip
		\subfloat[SURICATA\label{non-payload-direct-SURICATA}]{
			\protect\centering{}
			
			\begin{tabular}{>{\raggedleft}p{3.6cm}c>{\centering}p{1.2cm} r}
				
				\Xhline{2\arrayrulewidth} \noalign{\smallskip}
				
				\multirow{2}{3.6cm}{\textbf{~}} & \multicolumn{3}{c}{\textbf{Direct Attacks}}        
				\tabularnewline
				
				\expandafter\cline\expandafter{\expandafter2\string-4\smallskip} 
				& \textbf{Detected} & \textbf{Total} & \textbf{\%~~} \tabularnewline             
				
				\Xhline{2\arrayrulewidth} \noalign{\smallskip}
				
				\textbf{Apache Tomcat} & ~~~~~56 $+$5 & 61 & 100.00 \tabularnewline
				
				\textbf{DistCC} & 0 & 12 & 0.00 \tabularnewline
				
				\textbf{MSSQL} & 31 & 31 & 100.00 \tabularnewline
				
				\textbf{PostgreSQL} & 0 & 13 & 0.00 \tabularnewline             
				
				\textbf{Samba} & 0 & 19 & 0.00 \tabularnewline
				
				\textbf{Server} & 26 & 26 & 100.00 \tabularnewline                                       
				
				\Xhline{1\arrayrulewidth}\noalign{\smallskip}
				
				\textbf{Overall Detection} & 118  & 162 & 72.84 \tabularnewline
				\textbf{ADR$^*$ per Service} &  &  & 50.00 \tabularnewline
				\Xhline{2\arrayrulewidth} 
				\multicolumn{4}{r}{$^*$Average detection rate.}
			\end{tabular}       
			
		}
	\end{centering}  
	\caption{Detection of direct attacks in the ASNM-NPBO dataset by SNORT and SURICATA.}
	\label{non-payload-NIDS-Snort-Suricata-direct}
	
\end{table} \hspace{-0.5cm} \setlength{\tabcolsep}{1.4pt}
We can see in the tables that SNORT overcame SURICATA and correctly detected $100.00\%$ of direct attacks.
However, only 33 direct attacks on Apache service were detected by high priority rules of SNORT, and 24 attacks were undetected.
Despite it, we considered these attacks as correctly detected, as they occurred almost at the same time as other correctly predicted direct attacks, and thus might be a part of their execution.
In the case of SURICATA, the only one such undetected direct attack occurred.
Nevertheless, unlike SNORT, SURICATA did not fire any alert representing buffer overflow, shellcode, or remote command execution, but instead fired combination of high priority alerts 
related to potential corporate privacy violation:
\begin{itemize}
	\fontsize{8pt}{10pt}
	
	\item  
	\begingroup
	\begin{verbatim}
	ET POLICY 
	Incoming Basic Auth Base64 HTTP 
	Password detected unencrypted
	\end{verbatim}
	\endgroup
	\vspace{0.2cm}

	\item
	\begingroup
	\begin{verbatim}
	ET POLICY 
	Outgoing Basic Auth Base64 HTTP 
	Password detected unencrypted
	\end{verbatim}
	\endgroup
	\vspace{0.2cm}    
	
	\item    
	\begingroup
	\begin{verbatim}
	ET POLICY 
	HTTP Request on Unusual Port Possibly Hostile
	\end{verbatim}
	\endgroup
	\vspace{0.2cm}    
	
	\item
	\begingroup
	\begin{verbatim}
	ET POLICY 
	Internet Explorer 6 in use
	Significant Security Risk,
	\end{verbatim}
	\endgroup
\end{itemize}
which we decided to consider as correctly detected.
If we would not consider them as correctly detected, then SURICATA were not detected any attack on the Apache service.

\medskip \noindent
Next, we analyzed detection capabilities of both NIDSs on obfuscated attacks and the results are depicted in \autoref{non-payload-NIDS-Snort-Suricata-obfus}.
\setlength{\tabcolsep}{4pt}
\begin{table}[t]
	\centering{}       
	\smallskip\smallskip
	
	\begin{centering}        
		\subfloat[SNORT\label{non-payload-obfus-SNORT}]{
			\protect\centering{}
			
			\begin{tabular}{>{\raggedleft}p{3.6cm}c>{\centering}p{1.2cm} r}
				
				\Xhline{2\arrayrulewidth} \noalign{\smallskip}
				
				\multirow{2}{3.6cm}{\textbf{~}} & \multicolumn{3}{c}{\textbf{Obfuscated Attacks}}        
				\tabularnewline
				
				\expandafter\cline\expandafter{\expandafter2\string-4\smallskip} 
				& \textbf{Detected} & \textbf{Total} & \textbf{\%~~} \tabularnewline             
				
				\Xhline{2\arrayrulewidth} \noalign{\smallskip}
				
				\textbf{Apache Tomcat} & ~~~~~~128 $+$36 & 164 & 100.00 \tabularnewline
				
				\textbf{DistCC} & 23 & 23 & 100.00 \tabularnewline
				
				\textbf{MSSQL} & 103 & 103 & 100.00 \tabularnewline
				
				\textbf{PostgreSQL} & 45 & 45 & 100.00 \tabularnewline             
				
				\textbf{Samba} & 44 & 44 & 100.00 \tabularnewline
				
				\textbf{Server} & 98 & 100 & 98.00 \tabularnewline                                       
				
				\Xhline{1\arrayrulewidth}\noalign{\smallskip}
				
				\textbf{Overall Detection} & 478 & 480 & 99.58 \tabularnewline
				\textbf{ADR$^*$ per Service} &  &  & 99.67 \tabularnewline
				\Xhline{2\arrayrulewidth} 
				\multicolumn{4}{r}{$^*$Average detection rate.}
			\end{tabular}     
			
		}
	\end{centering}
	
	\begin{centering}
		\smallskip
		\subfloat[SURICATA\label{non-payload-obfus-SURICATA}]{
			\protect\centering{}
			
			\begin{tabular}{>{\raggedleft}p{3.6cm}c>{\centering}p{1.2cm} r}
				
				\Xhline{2\arrayrulewidth} \noalign{\smallskip}
				
				\multirow{2}{3.6cm}{\textbf{~}} & \multicolumn{3}{c}{\textbf{Obfuscated Attacks}}        
				\tabularnewline
				
				\expandafter\cline\expandafter{\expandafter2\string-4\smallskip} 
				& \textbf{Detected} & \textbf{Total} & \textbf{\%~~} \tabularnewline             
				
				\Xhline{2\arrayrulewidth} \noalign{\smallskip}
				
				\textbf{Apache Tomcat} & ~~~~~162 $+$1 & 163 & 100.00 \tabularnewline
				
				\textbf{DistCC} & 0 & 23 & 0.00 \tabularnewline
				
				\textbf{MSSQL} & 103 & 103 & 100.00 \tabularnewline
				
				\textbf{PostgreSQL} & 0 & 45 & 0.00 \tabularnewline             
				
				\textbf{Samba} & 0 & 44 & 0.00 \tabularnewline
				
				\textbf{Server} & 98 & 100 & 98.00 \tabularnewline                                       
				
				\Xhline{1\arrayrulewidth}\noalign{\smallskip}
				
				\textbf{Overall Detection} & 364 & 478 & 76.15 \tabularnewline
				\textbf{ADR$^*$ per Service} &  &  & 49.67 \tabularnewline
				\Xhline{2\arrayrulewidth} 
				\multicolumn{4}{r}{$^*$Average detection rate.}
			\end{tabular}              
			
		}
	\end{centering}  
	\caption{Detection of obfuscated attacks in ASNM-NPBO dataset by SNORT and SURICATA.}
	\label{non-payload-NIDS-Snort-Suricata-obfus}
	
\end{table} \hspace{-0.5cm} \setlength{\tabcolsep}{1.4pt}
Comparing the detection rate of SNORT and SURICATA on obfuscated attacks, we can conclude that SNORT overcame SURICATA again and the ratio of their correct detection was almost the same as in the case of direct attacks (see \autoref{non-payload-NIDS-Snort-Suricata-direct}).
The only difference occurred during the exploitation of a vulnerability in Server service, where two instances of obfuscated attacks were not detected by any NIDS.
These two instances utilized obfuscations with IDs (f) and (g), both from a category of unreliable network traffic channel simulation techniques (see  \autoref{tab:NPBO-obfuscation-techniques}). 
There were also several undetected obfuscated attacks on Apache service in both NIDSs, but we were able to track their occurrences and associate them as part of other correctly detected attacks; hence, the detection rate for Apache service achieved 100.00\% for both NIDSs. 
Regarding Apache service, SURICATA once again did not fire any alert detecting malicious content, but instead, it fired the previously mentioned combination of high priority alerts stating corporate privacy violation, which we, once again, considered as a correct detection.
Also, note that SURICATA fired one non-high-priority alert classified as potentially bad traffic in all instances of direct and obfuscated attacks exploiting PostgreSQL service:
\begin{itemize}
	\fontsize{8pt}{10pt}
	
	\item
	\begin{verbatim}
	ET POLICY
	Suspicious inbound to PostgreSQL port 5432.    
	\end{verbatim}
\end{itemize}
However, we did not consider it as a correct detection due to the low priority of the alert.
As discussed in \autoref{sec:desc-TUN}, VirusTotal likely uses a paranoid rule set, and thus fired alerts may contain false positives.
Comparing fired alerts before and after obfuscation, we can see that utilized NIDSs detected most of the obfuscated attacks by non-payload-based, but there were also a few cases where they failed, and thus, evasion was successful. 
\vspace{0.3cm}
\section{Benchmarking the Datasets}\label{sec:benchmarking}
In the previous research~\cite{homoliakasnm2013,2016-ihomoliak-thesis,homoliak:NBAofObfNetVul,homoliak:ChofBOAinHTTP,homoliak2016exploitation,ih2019nonpayload}, we conducted several machine learning experiments with ASNM datasets, and we summarize them in the current section.

\subsection{ASNM-CDX-2009 Dataset}
\paragraph{Forward Feature Selection}
First, we used 5-fold cross-validation and forward feature selection (FFS) on top of the Naive Bayes classifier with kernel functions for the estimation of density distribution, which represents a non-parametric estimation method. 
In FFS, we accepted one iteration without improvement as we wanted to avoid the selection process to get stuck in local extremes. 
The maximal number of selected features was limited to 20 (although it was never reached).
We used the binary label of the dataset (i.e., $label\_2$), and we obtained $\varDash{F_{1}-measure}$ over $90\%$ and an average recall of both classes equal to $92\%$. 

Additionally, we compared the performance of ASNM features with discriminators of A. Moore~\cite{moore2005discriminators} in the same setting, and we concluded that both feature sets yielded similar results. 
Moreover, when we merged both feature sets and rerun FFS, $\varDash{F_{1}-measure}$ reached $98.87\%$~\cite{2016-ihomoliak-thesis}.
\begin{figure}[hb]
    \begin{centering}
        \includegraphics[width=0.98\columnwidth]{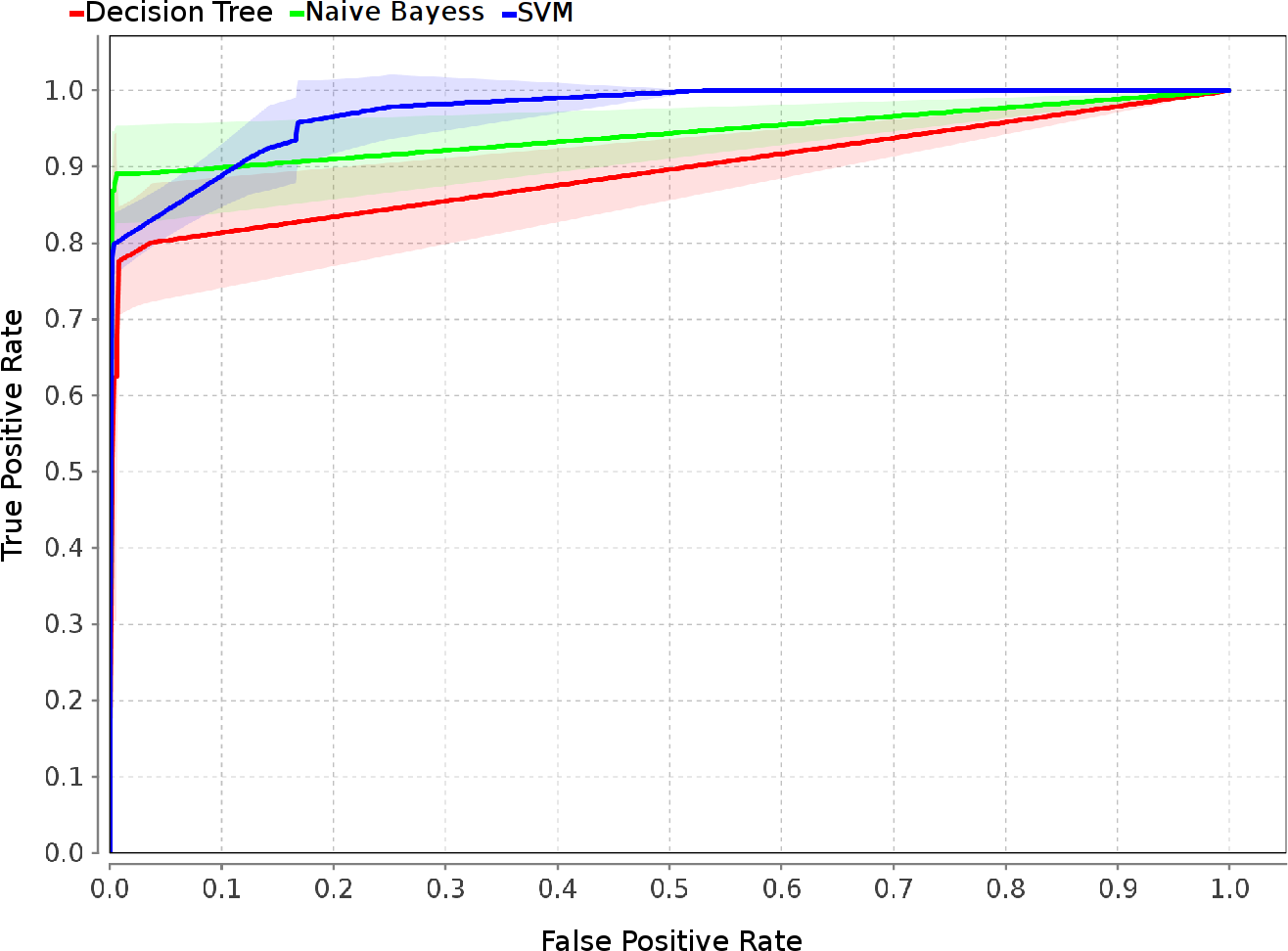}
        \par\end{centering}
    \caption{ROC diagram comparing a few classifiers on the ASNM-CDX-2009 dataset.}
    \label{fig:cdx:neutral-ROC}
\end{figure}

\paragraph{Comparison of Several Classifiers}
Next, we compared three non-parametric classifiers while using a subset of ASNM features obtained by FFS with the Naive Bayes classifier -- the selected features are enumerated and described in \autoref{FFS-features-CDX} of Appendix.
The individual confusion matrices that we obtained are presented in \autoref{tab:performance_AIPS:CDX_on_ASNM} (Naive Bayes with kernel density estimation), \autoref{performance_AIPS:CDX_on_asnm_decision_tree} (Decision Tree with Gini index as a selection criterion
for splitting of attributes), and \autoref{performance_AIPS:CDX_on_asnm_SVM} (SVM with radial basis function).
Finally, all classifiers were compared using ROC curves, and a comparison is depicted in \autoref{fig:cdx:neutral-ROC}.
Note that ROC curves also depict a variance coming from a cross-validation method, which is shown by line-adjacent transparent areas.

\setlength{\tabcolsep}{4pt} 
\begin{table}[t]       
    \centering{}
    \begin{tabular}{>{\raggedright}p{1.3cm}ccc>{\centering}p{1.8cm}}
        \toprule
        
        \multicolumn{2}{c}{\textbf{Classification Accuracy:}} & \multicolumn{2}{c}{\textbf{True Class}} & \multirow{2}{2cm}{\centering{}\textbf{Precision}} \tabularnewline
        \expandafter\cline \expandafter{\expandafter3\string-4\smallskip}          
        
        \multicolumn{2}{c}{99.86\% $\pm 0.07$} & \textbf{Legit. Flows} & \textbf{Attacks} & \tabularnewline
        \noalign{\smallskip} \Xhline{2\arrayrulewidth} \noalign{\smallskip}
        
        \multirow{2}{1cm}{\textbf{\specialcell{Predicted\\~~~Class}}} & \textbf{Legit. Flows} & 5726 & 7 & 99.88\%\tabularnewline
        
        & \textbf{~~~~~~Attacks} & 1 & 37 & 97.37\%\tabularnewline
        \midrule 
        \multicolumn{2}{r}{\textbf{Recall}} & 99.98\% & 84.09\% & $ F_1 = $ 90.24\% \tabularnewline
        \bottomrule
    \end{tabular}    
    \caption{Performance of the Naive Bayes classifier on the ASNM-CDX-2009 dataset.}
    \label{tab:performance_AIPS:CDX_on_ASNM}
    
    \medskip
    \begin{tabular}{>{\raggedright}p{1.3cm}ccc>{\centering}p{1.8cm}}
        \Xhline{2\arrayrulewidth} \noalign{\smallskip} 
        
        \multicolumn{2}{c}{\textbf{Classification Accuracy:}} & \multicolumn{2}{c}{\textbf{True Class}} & \multirow{2}{2cm}{\centering{}\textbf{Precision}} \tabularnewline
        \expandafter\cline \expandafter{\expandafter3\string-4\smallskip}          
        
        \multicolumn{2}{c}{99.71\% $\pm 0.07$} & \textbf{Legit. Flows} & \textbf{Attacks} & \tabularnewline
        \noalign{\smallskip} \Xhline{2\arrayrulewidth} \noalign{\smallskip}
        
        \multirow{2}{1cm}{\textbf{\specialcell{Predicted\\~~~Class}}} & \textbf{Legit. Flows} & 5721 & 11 & 99.81\%\tabularnewline
        
        & \textbf{~~~~~~Attacks} & 6 & 33 & 84.62\%\tabularnewline
        \midrule
        \multicolumn{2}{r}{\textbf{Recall}} & 99.90\% & 75.00\% & $ F_1 = $ 79.52\% \tabularnewline
        \Xhline{2\arrayrulewidth}
    \end{tabular}    
    \caption{Performance of the decision tree classifier on the ASNM-CDX-2009 dataset.}
    \label{performance_AIPS:CDX_on_asnm_decision_tree}
    
    \medskip
    \begin{tabular}{>{\raggedright}p{1.3cm}ccc>{\centering}p{1.8cm}}
        \Xhline{2\arrayrulewidth} \noalign{\smallskip} 
        
        \multicolumn{2}{c}{\textbf{Classification Accuracy:}} & \multicolumn{2}{c}{\textbf{True Class}} & \multirow{2}{2cm}{\centering{}\textbf{Precision}} \tabularnewline
        \expandafter\cline \expandafter{\expandafter3\string-4\smallskip}          
        
        \multicolumn{2}{c}{99.81\% $\pm 0.06$} & \textbf{Legit. Flows} & \textbf{Attacks} & \tabularnewline
        \noalign{\smallskip} \Xhline{2\arrayrulewidth} \noalign{\smallskip}
        
        \multirow{2}{1cm}{\textbf{\specialcell{Predicted\\~~~Class}}} & \textbf{Legit. Flows} & 5726 & 10 & 99.83\%\tabularnewline
        
        & \textbf{~~~~~~Attacks} & 1 & 34 & 97.4\%\tabularnewline
        \midrule 
        \multicolumn{2}{r}{\textbf{Recall}} & 99.98\% & 77.27\% & $ F_1 = $ 86.07\% \tabularnewline
        \Xhline{2\arrayrulewidth}
    \end{tabular}    
    \caption{Performance of the SVM classifier on the ASNM-CDX-2009 dataset.}
    \label{performance_AIPS:CDX_on_asnm_SVM}
    \vspace{-0.4cm}
\end{table}

\subsection{ASNM-TUN Dataset}\label{sec:TUN-eval}

\paragraph{Forward Feature Selection}
Alike the case of the previous dataset, we again started with the FFS method using the same Naive Bayes classifier and 5-fold cross-validation, while we allowed acceptance of one FFS iteration without improvement to avoid the selection process becoming stuck in local extremes. 
All cross-validation experiments have been adjusted to employ stratified sampling during assembling of folds, which ensured equally balanced class distribution of each fold. 
We performed two-class prediction (i.e., using the label denoted as $label\_2$).
Some features existed, which were inconvenient for comparison of synthetic attacks with legitimate traffic captured in a real network; therefore, such features were removed from the dataset in the pre-processing phase of our experiment.
The examples include TTL-based features, IP addresses, ports, MAC addresses, the occurrence of source/destination host in the monitored local network, some context-based features, etc.
The experiment consisted of two executions of FFS. 
The first took as an input just legitimate traffic and direct attack entries and represented the case where the classifier was trained without knowledge about obfuscated attacks.
The second execution took as input the whole dataset of network traffic -- consisting of legitimate traffic, direct attacks as well as obfuscated ones, and therefore represented the case where the classifier was aware of obfuscated attacks.
The selected features of both executions are depicted in \autoref{FFS-features-tunneling} of Appendix. 
The penultimate column of the table (i.e., FFS DOL) denotes the selected features where the whole dataset was utilized for the FFS, and the last column (i.e., FFS DL) denotes the case where only direct attacks and legitimate traffic were taken into account.

Several mutual features were selected in both cases, which means they provided a~value regardless of whether obfuscation was performed or not.
Almost all of the following experiments will use the feature set gained from the second execution (i.e., FFS DOL), as we consider them as more appropriate for general behavior representation of both kinds of attacks.

\paragraph{Evasions}
First, we executed an experiment that performed detection of malicious obfuscated attacks by the classifier trained on all direct attacks and legitimate traffic samples.
It represented the situation when the classifier had no previous knowledge about obfuscated attacks, and therefore we used FFS DL feature set.
As a result, only $35.63\%$ of obfuscated attacks (i.e., $31$ of $87$) were correctly detected by the classifier, and thus an average recall and $\varDash{F_{1}-measure}$ of the classifier were equal to $67.53\%$ and $52.10\%$, respectively.  
An associated confusion matrix is depicted in \autoref{wholeDataPredictionTunneling}.
\setlength{\tabcolsep}{4pt}
\begin{table}[th]       
    \centering{}
    
    \begin{tabular}{>{\raggedright}p{1.6cm}ccc>{\centering}p{1.8cm}}
        \toprule \noalign{\smallskip} 
        
        \multicolumn{2}{c}{\textbf{Classification Accuracy:}} & \multicolumn{2}{c}{\textbf{True Class}} & \multirow{2}{2cm}{\centering{}\textbf{Precision}} \tabularnewline
        \expandafter\cline \expandafter{\expandafter3\string-4\smallskip}          
        
        \multicolumn{2}{c}{78.41\%} & \textbf{\specialcell{Legit.\\Flows}} & \textbf{\specialcell{~Obfus.\\Attacks}} & \tabularnewline
        \noalign{\smallskip} \toprule \noalign{\smallskip}
        
        \multirow{2}{1cm}{\textbf{Predicted Class}} & \textbf{Legit. Flows} & 176 & 56 & 75.86\%\tabularnewline
        
        & \textbf{~Obfus. Attacks} & 1 & 31 & 96.88\%\tabularnewline
        \midrule 
        \multicolumn{2}{r}{\textbf{Recall}} & 99.44\% & 35.63\% & $ F_1 = $ 52.10\% \tabularnewline
        \bottomrule
    \end{tabular}    
    
    \caption{Detection of unknown obfuscated attacks by the Naive Bayes classifier trained on all direct attack samples and legitimate traffic samples from the ASNM-TUN dataset.}
    \label{wholeDataPredictionTunneling}
\end{table} \hspace{-0.5cm} \setlength{\tabcolsep}{1.4pt} 
We realized that $64.36\%$ of obfuscated attacks were incorrectly predicted as legitimate traffic, and thus caused an evasion of the classifier.

\paragraph{Training Data Augmentation}
Our second binary classification experiment considered explicit information about obfuscated attacks in the training phase of the classifier.
Therefore, we used direct and obfuscated attacks labeled as one class while using 5-fold cross-validation. 
FFS DOL feature set was used for the purpose of this experiment.
The resulting confusion matrix with performance measures is shown in \autoref{referenceBinaryTunneling}.
The outcome of this experiment indicates a~high class recall and $\varDash{F_{1}-measure}$ of the classifier trained with knowledge about some obfuscated attacks.  
\setlength{\tabcolsep}{4pt}
\begin{table}[b]       
    \centering{}
    
    \begin{tabular}{>{\raggedright}p{1.6cm}ccc>{\centering}p{1.8cm}}
        \toprule \noalign{\smallskip} 
        
        \multicolumn{2}{c}{\textbf{Classification Accuracy:}} & \multicolumn{2}{c}{\textbf{True Class}} & \multirow{2}{2cm}{\centering{}\textbf{Precision}} \tabularnewline
        \expandafter\cline \expandafter{\expandafter3\string-4\smallskip}          
        
        \multicolumn{2}{c}{99.49\% $\pm$ 0.62\%} & \textbf{\specialcell{Legit.\\Flows}} & \textbf{\specialcell{~~All\\Attacks}}  & \tabularnewline
        \noalign{\smallskip} \toprule \noalign{\smallskip}
        
        \multirow{2}{1cm}{\textbf{Predicted Class}} & \textbf{Legit. Flows} & 176 & 1 & 99.44\%\tabularnewline
        
        & \textbf{~All Attacks} & 1 & 216 & 99.54\%\tabularnewline
        \midrule 
        \multicolumn{2}{r}{\textbf{Recall}} & 99.44\% & 99.54\% & $ F_1 = $ 99.54\% \tabularnewline
        \bottomrule
    \end{tabular}    
    
    \caption{Performance of the Naive Bayes classifier on the ASNM-TUN dataset using the binary label (i.e., $label\_2$).}
    \label{referenceBinaryTunneling}
\end{table} \hspace{-0.5cm} \setlength{\tabcolsep}{1.4pt}

\paragraph{Comparison of Several Classifiers}
For the purpose of performance comparison of various classifiers, we executed 5-fold cross-validation on the other two non-parametric classifiers -- decision tree and SVM. 
FFS DOL feature set was used in this experiment as the input for the classifiers working with two class prediction (i.e., using $label\_2$).
At first, we evaluated the performance of the SVM classifier, which utilized a radial basis function as the non-linear kernel. 
The adjacent confusion matrix is depicted in \autoref{tunneling:svm_binominal}.
\setlength{\tabcolsep}{4pt}
\begin{table}[th]       
    \centering
    
    \begin{tabular}{>{\raggedright}p{1.6cm}ccc>{\centering}p{1.8cm}}
        \toprule \noalign{\smallskip} 
        
        \multicolumn{2}{c}{\textbf{Classification Accuracy:}} & \multicolumn{2}{c}{\textbf{True Class}} & \multirow{2}{2cm}{\centering{}\textbf{Precision}} \tabularnewline
        \expandafter\cline \expandafter{\expandafter3\string-4\smallskip}          
        
        \multicolumn{2}{c}{80.96\% $\pm$ 3.51\%} & \textbf{\specialcell{Legit.\\Flows}} & \textbf{\specialcell{~~All\\Attacks}} & \tabularnewline
        \noalign{\smallskip} \toprule \noalign{\smallskip}
        
        \multirow{2}{1cm}{\textbf{Predicted Class}} & \textbf{Legit. Flows} & 176 & 74 & 70.40\%\tabularnewline
        
        & \textbf{~All Attacks} & 1 & 143 & 99.31\%\tabularnewline
        \midrule 
        \multicolumn{2}{r}{\textbf{Recall}} & 99.44\% & 65.90\% & $ F_1 = $ 79.22\% \tabularnewline
        \bottomrule
    \end{tabular}        
    \caption{Performance of the SVM classifier on the ASNM-TUN dataset.}
    \label{tunneling:svm_binominal}
    
    \medskip
    \begin{tabular}{>{\raggedright}p{1.6cm}ccc>{\centering}p{1.8cm}}
        \toprule \noalign{\smallskip} 
        
        \multicolumn{2}{c}{\textbf{Classification Accuracy:}} & \multicolumn{2}{c}{\textbf{True Class}} & \multirow{2}{2cm}{\centering{}\textbf{Precision}} \tabularnewline
        \expandafter\cline \expandafter{\expandafter3\string-4\smallskip}          
        
        \multicolumn{2}{c}{95.93\% $\pm$ 2.47\%} & \textbf{\specialcell{Legit.\\Flows}} & \textbf{\specialcell{~~All\\Attacks}} & \tabularnewline
        \noalign{\smallskip} \toprule \noalign{\smallskip}
        
        \multirow{2}{1cm}{\textbf{Predicted Class}} & \textbf{Legit. Flows} & 169 & 8 & 95.48\%\tabularnewline
        
        & \textbf{~All Attacks} & 8 & 209 & 96.31\%\tabularnewline
        \midrule 
        \multicolumn{2}{r}{\textbf{Recall}} & 95.48\% & 96.31\% & $ F_1 = $ 96.31\% \tabularnewline
        \bottomrule
    \end{tabular}    
    
    \caption{Performance of the decision tree classifier on the ASNM-TUN dataset.}
    \label{tunneling:decision_tree_binominal}

\end{table} \hspace{-0.5cm} \setlength{\tabcolsep}{1.4pt} 
The next experiment was performed with the decision tree classifier, which utilized gini index as selection criterion for splitting of attributes.
The adjacent result is represented by confusion matrix in \autoref{tunneling:decision_tree_binominal}.
The results of both performance evaluation experiments can be compared to the result of the Naive Bayes classifier represented by the confusion matrix from  \autoref{referenceBinaryTunneling}.
Considering an average recall of all classes and $\varDash{F_{1}-measure}$, we can say that the Naive Bayes classifier achieved the best results, following by the decision tree, and finally by SVM.

\begin{figure}[t]
    \begin{centering}
        \includegraphics[width=0.98\columnwidth]{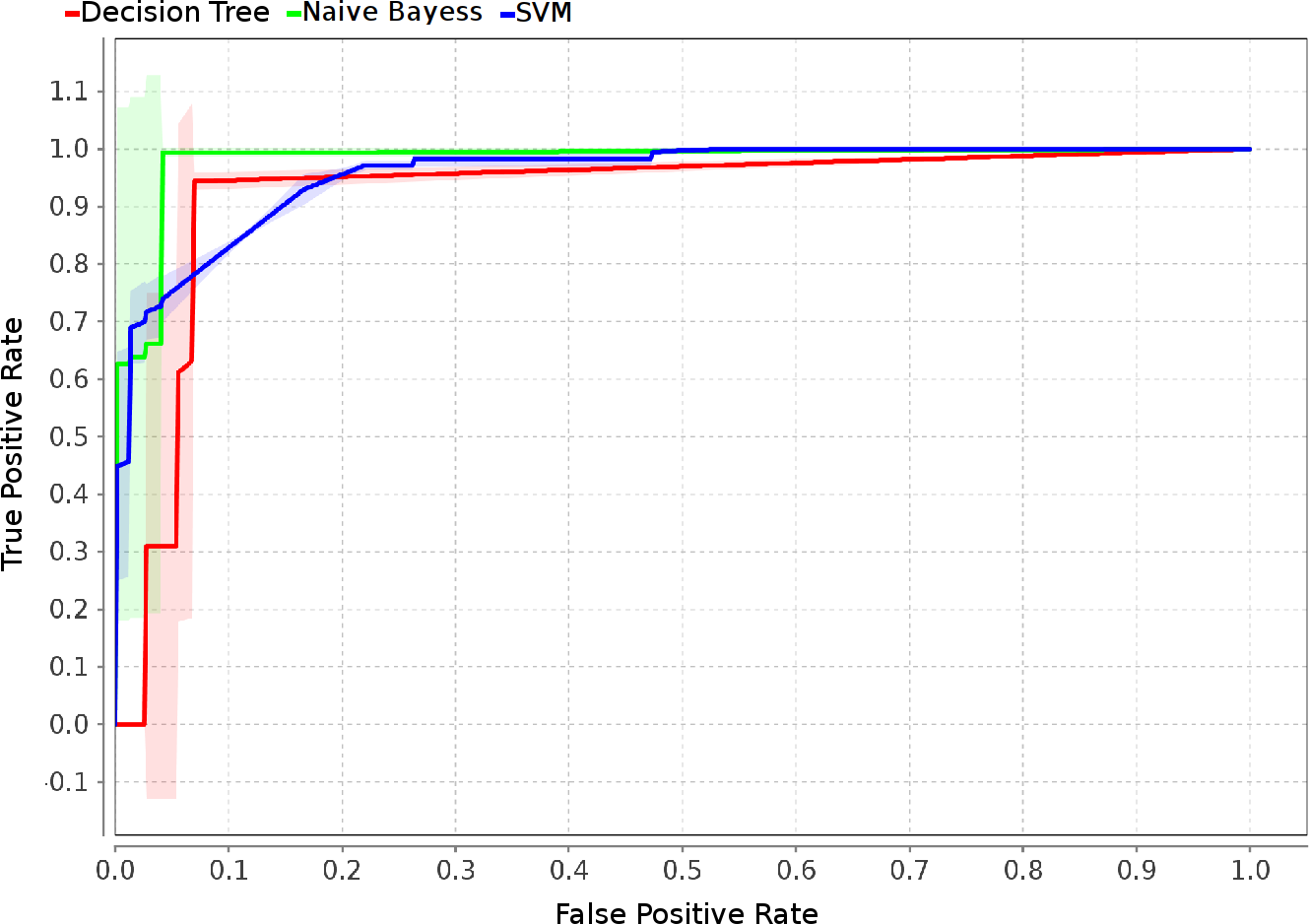}
        \par\end{centering}
    \caption{ROC diagram comparing a few classifiers on the ASNM-TUN dataset.}
    \label{fig:tunneling:neutral-ROC}
\end{figure}

All the classification models were compared by ROC method (see \autoref{fig:tunneling:neutral-ROC}). 
Note that comparison of ROC ran above the cross-validation method, and thus generated certain variability, which is shown by line-adjacent transparent areas.
 \hspace{-0.5cm}
For more experiments with this dataset, including tri-nominal and multi-nominal labels and individual feature analysis, we refer the reader to~\cite{homoliak:NBAofObfNetVul}, ~\cite{homoliak:ChofBOAinHTTP}, and~\cite{2016-ihomoliak-thesis}.

\subsection{ASNM-NPBO Dataset}

\paragraph{Forward Feature Selection}
Alike the case of the previous datasets, we again started with the FFS method using the same Naive Bayes classifier and 5-fold cross-validation, while we allowed acceptance of one FFS iteration without improvement, and we excluded the same inconvenient features as in \autoref{sec:TUN-eval}.
We performed two-class prediction (i.e., using $label\_2$) in two executions of FFS using the Naive Bayes classifier -- the first execution did not contain obfuscated attack samples (i.e., FFS DL) and the another one included these samples (i.e., FFS DOL).
The selected features of both executions are depicted in \autoref{FFS-features-NPBO} of Appendix.

\paragraph{Evasions}
5-fold cross-validation with FFS DL features was performed using all direct attack samples and legitimate traffic samples.
The performance measures of three classifiers validated by the cross-validation are shown in \autoref{tab:evasion-of-classifiers}.
Then the classifiers trained on all direct attacks and legitimate traffic samples were applied for the prediction of the obfuscated attacks and all attacks, respectively (see \autoref{tab:prediction-obfus-and-all}).\footnote{Note that we do not depict FPRs in the tables since no changes to legitimate traffic was made, hence FPRs remain the same as in \autoref{tab:evasion-of-classifiers}.} 
Here TPRs were deteriorated for all classifiers, which means that some obfuscated attacks were successful -- they were predicted as legitimate traffic, and thus caused evasion of the classifiers. 
\setlength{\tabcolsep}{4pt}
\begin{table}[t]
    \centering
    \begin{footnotesize}        
        \begin{tabular}{r r r r r}
            \toprule
            \textbf{Classifier} & \textbf{TPR} & \textbf{FPR} & $\mathbf{F_1 ~(\uparrow)}$~~ & \textbf{Avg. Recall}  \\
            \midrule
            
            \textbf{Naive Bayes} & 98.15\% & 0.02\% & 98.45\% & 99.07\%  \\
            
            \textbf{Decision Tree} & 95.68\% & 0.09\% & 94.80\% & 97.80\% \\                        
            
            \textbf{SVM}  & 82.72\% & 0.01\% & 90.24\% & 91.36\% \\

            \bottomrule
        \end{tabular}
    \end{footnotesize}
    \smallskip
    \caption{Direct attacks and legitimate traffic cross validation on ASNM-NPBO dataset.}\label{tab:evasion-of-classifiers}
\end{table}    \setlength{\tabcolsep}{1.4pt}    
\setlength{\tabcolsep}{4pt}
\begin{table}[t]
    \centering
    \begin{footnotesize}        
        \subfloat[\label{tab:prediction-obfus}Obfuscated attacks]{
            \begin{tabular}{r r r}
                \toprule
                \textbf{Classifier} & \textbf{TPR $(\uparrow)$} & \textbf{$\mathbf{\Delta}$ TPR} \\
                \midrule

                \textbf{Naive Bayes}  & 52.30\% & -45.85\% \\
                
                \textbf{Decision Tree} & 36.61\% & -59.07\% \\
                
                \textbf{SVM}  &  15.90\% & -66.82\% \\

                \bottomrule
            \end{tabular}
        }\\
        \subfloat[\label{tab:prediction-all}All attacks]{
            \begin{tabular}{r r r}
                \toprule
                \textbf{Classifier} & \textbf{TPR $(\uparrow)$} & \textbf{$\mathbf{\Delta}$ TPR} \\
                \midrule

                \textbf{Naive Bayes}  & 64.38\% & -33.77\%\\
                
                \textbf{Decision Tree} &  52.03\% & -43.65\% \\
                
                \textbf{SVM}  & 26.25\% & -56.47\% \\

                \bottomrule
            \end{tabular}
        }
    \end{footnotesize}
    \caption{Prediction of obfuscated attacks and all attacks in the ASNM-NPBO dataset by classifiers trained without knowledge about obfuscated attacks.}
    \label{tab:prediction-obfus-and-all}
\end{table}    \setlength{\tabcolsep}{1.4pt}

\paragraph{Training Data Augmentation}
To improve the resistance of the classifiers against evasions, we widened their knowledge about different mixtures of obfuscated attack instances,
which was accomplished by random 5-fold cross-validation of the whole dataset. 
In this experiment, we use FFS DOL features that consider knowledge about obfuscated attacks for updating not only the model of the classifier but also the underlying feature set (in contrast to the previous experiment). 
Additionally, we show the results with FFS DL features, which consider updating the model only.
The results of this experiment are shown in \autoref{tab:bi-all-cross}.
Comparing against the results from the previous experiment (see FPRs from \autoref{tab:evasion-of-classifiers} and TPRs from \autoref{tab:prediction-obfus-and-all}b), 
most of the classifiers were significantly improved in TPR, while FPR was deteriorated only slightly.
Hence, the classifiers trained with knowledge about some obfuscated attacks were able to detect the same and similar obfuscated attacks later.

\setlength{\tabcolsep}{3pt}
\begin{table}
	\centering
	\vspace{0.2cm}
	\begin{footnotesize}
		
		\subfloat[\label{tab:prediction-obfus-ffs-dl}FFS DL features]{
			\begin{centering}
				\protect\centering{}
				\begin{tabular}{r r r r r r r}
					\toprule
					\textbf{Classifier} & \textbf{TPR}~ & \textbf{FPR}~ & \specialcell{~~$\Delta$\\\textbf{TPR}} & \specialcell{~~$\Delta$\\ \textbf{FPR}} & $\mathbf{F_1} (\uparrow)$~~~ & \textbf{\specialcell{~Avg.\\Recall}} \\
					\midrule
					
					\textbf{Naive Bayes}  & 93.28\% & 0.73\% & +28.90\% &    +0.71\%  & 90.73\% & 96.28\%  \\
					
					\textbf{SVM}  &  80.31\% & 0.05\% & +54.06\% & +0.04\% & 88.70\% & 90.13\% \\

					\textbf{Decision Tree} & 67.34\% & 0.36\% &  +15.31\% &    +0.27\% & 77.65\% & 83.49\% \\

					\bottomrule
				\end{tabular}\protect
			\end{centering}
		}
		
		\subfloat[\label{tab:prediction-obfus-ffs-dol}FFS DOL features]{
			\begin{centering}
				
				\protect\centering{}
				\begin{tabular}{r r r r r r r}
					\toprule
					\textbf{Classifier} & \textbf{TPR}~ & \textbf{FPR}~ & \specialcell{~~$\Delta$\\\textbf{TPR}} & \specialcell{~~$\Delta$\\ \textbf{FPR}} & $\mathbf{F_1} (\uparrow)$~~~ & \textbf{\specialcell{~Avg.\\Recall}} \\
					\midrule
					
					\textbf{SVM} & 99.53\% & 0.13\% & +73.28\% & +0.12\% & 98.68\% & 99.70\% \\
					
					\textbf{Decision Tree} & 98.44\%    & 0.19\% & +46.41\% & +0.10\% & 97.60\% & 99.13\% \\
					
					\textbf{Naive Bayes} & 98.75\% & 0.99\% & +34.37\% & +0.97\% & 91.66\%    & 98.88\% \\

					\bottomrule
				\end{tabular}\protect
			\end{centering}
		}
		\caption{Cross validation of the whole ASNM-NPBO dataset, representing the situation where classifiers were aware of some obfuscated attacks, and therefore they brought performance improvement in contrast to classifiers aware only about direct attacks (see \autoref{tab:prediction-obfus-and-all}).}
		
		\label{tab:bi-all-cross}
	\end{footnotesize}
\end{table}    
\setlength{\tabcolsep}{1.4pt}

\paragraph{Comparison of Several Classifiers}
From the previous experiments, we can say that the Naive Bayes classifier was the least sensitive to evasions by non-payload-based obfuscations (see \autoref{tab:evasion-of-classifiers}), while SVM was the most sensitive classifier.
This might be caused by overfitting of the training data. 
Note that all classifiers used the feature sets selected by FFS with the Naive Bayes classifier. 
However, we also rerun FFS with individual classifiers, but obtained results were much worse than using the features selected by the Naive Bayes classifier.

After augmentation of a training data without updating the feature set (see \autoref{tab:bi-all-cross}a), we observe that the Naive Bayes classifier is the most robust one.
However, when making a training data augmentation with updating the feature set (see \autoref{tab:bi-all-cross}b), other classifiers perform better than Naive Bayes, which might be again caused by overfitting of them.

\begin{figure}
	\begin{centering}
		\includegraphics[width=0.98\columnwidth]{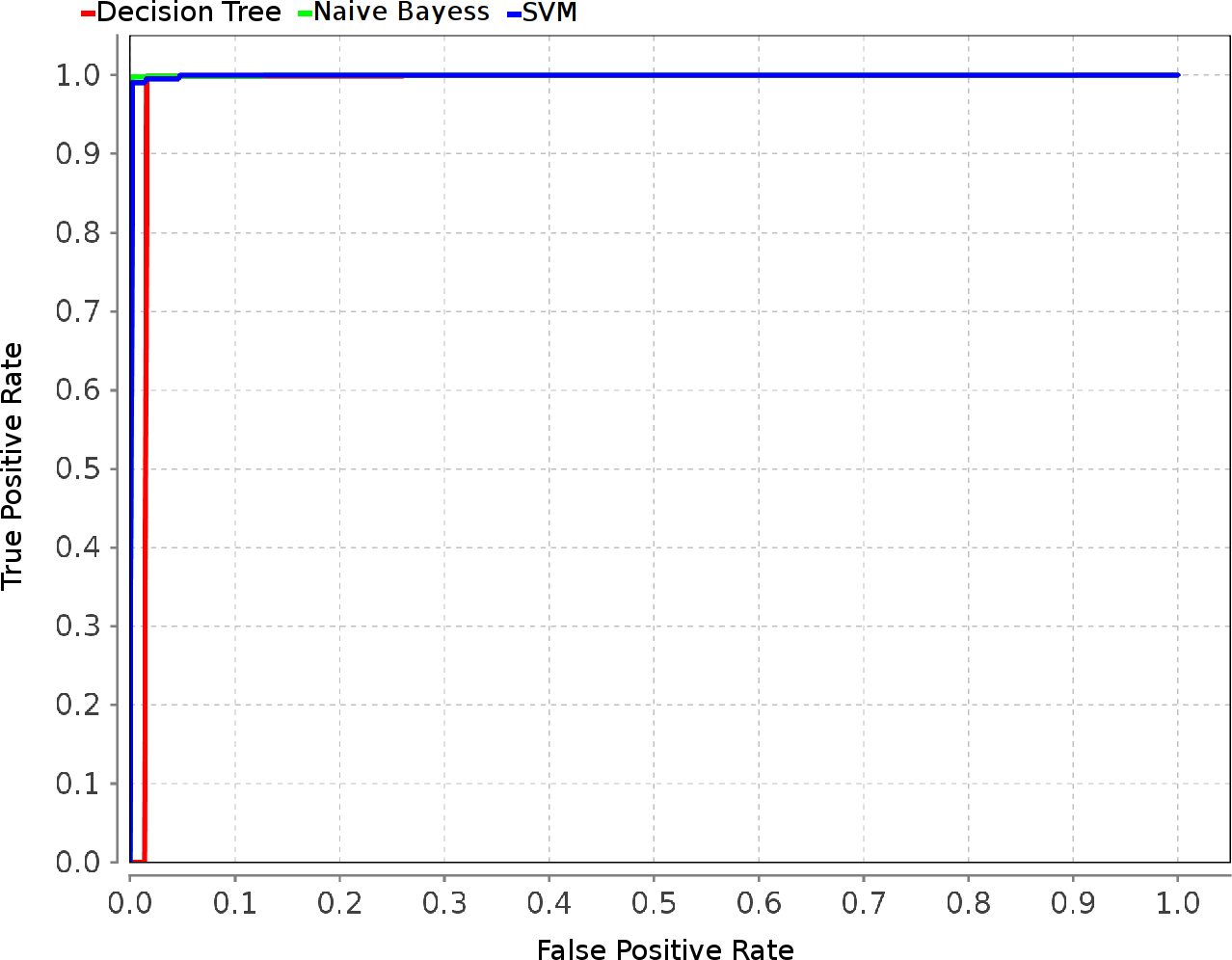}
		\par\end{centering}
	\caption{ROC diagram comparing a few classifiers on the ASNM-NPBO dataset.}
	\label{fig:non-pyload:neutral-ROC}
	\vspace{0.2cm}
\end{figure}

Finally, we compared the classification models by ROC method (see \autoref{fig:non-pyload:neutral-ROC}).
The best results were achieved in the case of the Naive Bayes classifier and SVM.
For more experiments with this dataset, including tri-nominal and multi-nominal labels, detection of unknown obfuscations by a custom leave-one-out validation, and individual feature analysis, we refer the reader to~\cite{ih2019nonpayload} and~\cite{2016-ihomoliak-thesis}.
 
\vspace{0.3cm}
\section{Related Work}\label{sec:related-work}
In this section, we summarize public datasets intended for the evaluation of network intrusion detection solutions. 
We partition all datasets into two categories. 
The first category represents datasets containing \textit{raw network traffic traces}, and the second category represents datasets containing \textit{high-level features} extracted from underlying network traces.

\subsection{Datasets of Network Traffic Traces}\label{SOA:DatasetsNetworkDumps}
Datasets from this category have one property in
common: they contain network traffic traces with optional data serving
for labeling purposes. 
The first four representatives of this category are large
collections of datasets and are referred to as projects -- MWS~\cite{hatada2015empowering}, PREDICT~\cite{PREDICTdataset},
CAIDA~\cite{caidaProject}, NETRESEC ~\cite{netresecDatasets}. 
The next four examples of this category
represent just one specific collection of network data and are referred to
as datasets -- DARPA~\cite{darpa-Url}, CCRC~\cite{massicotte2006automatic}, CDX~\cite{sangster2009toward}, CONTIAGO~\cite{ContagioMalwareDump}.
We describe them in the following.

\medskip
\subsubsection{\textbf{Project MWS}}\label{sub:MWS-Datasets}
The project MWS represents a~collection of various types of datasets that
are primarily intended for use in anti-malware research~\cite{hatada2015empowering}, but some of them are also applicable in network intrusion detection.
A~summary of the MWS datasets is available in Japanese
\cite{akiyama2014datasets,hatada2011datasets,hatada2010datasets,hatada2009dataset,kamizono2013datasets}, and it covers three categories of datasets, which is based on phases of attacks: 
(1) probing, 
(2) infection, and
(3) malware activities after infection.

From the perspective of network intrusion detection, we consider
PRACTICE, D3M, CCC, and NICTER as related datasets of MWS. 
However, for network intrusion detection is also important to have a ground truth, which can be inferred from the MWS datasets called FFRI, IIJ MITF, D3M, and CCC.
In the following, we briefly describe these datasets.

\smallskip
Cyber Clean Center (\textbf{CCC}) dataset consists of a~malware sample,
honeypot packet trace, and malware collection log. The dataset was
collected from server-side, high-interaction honeypots operated by
the CCC in a~distributed manner. Over a~hundred of honeypots gathered
attacks and collected malware through multiple ISPs. These honeypots
were based on Windows 2000 and Windows XP SP1 virtual machines.
Drive-by Download Data by Marionette (\textbf{D3M}) dataset is a~set
of packet traces collected from the web-client-based high-interaction
honeypot system Marionette~\cite{akiyama2010design,akiyama2015client},
which is built upon Internet Explorer with several vulnerable plugins,
such as Adobe Reader, Flash Player, WinZip, QuickTime. The
datasets contain packet traces for the two periods: at infection and
after infection.
The \textbf{IIJ MITF} dataset is collected by server-side, low interaction
honeypots based on the open-source honeypot Dionaea~\cite{DionaeaHoneypot}.
This dataset contains attack communication and malware collection
logs from a~hundred honeypots between July 2011 and April 2012 in
order to discover the trends of bot and botnets. 
The \textbf{PRACTICE} dataset contains
the packet traces obtained during long-term dynamic analysis of five
malware samples (Zbot, SpyEye, etc.) and their metadata, while the focus was put on network activity of malware, using the dynamic analysis system proposed in~\cite{aoki2011controlling}.
The analysis period of the dataset is one week in the middle of May 2013.
The \textbf{FFRI} dataset focuses on the internal activities that
occurred at a~host by the influence of malware and are generated by dynamic
analysis systems -- Cuckoo sandbox~\cite{guarnieri2012cuckoo} and
FFR Yarai Analyzer Professional~\cite{FFRyaraiAnalyzer}. 
The \textbf{NICTER} darknet dataset is a~set of packet traces collected from April 2011 to 2014 using the darknet monitoring system NICTER~\cite{inoue2008nicter}. The packet traces contain scan packets to
explore the reachable hosts by worms, backscatter
packets caused by source IP address spoofing, distributed reflection
denial of service (DRDoS) attacks using DNS and NTP, etc.

\medskip
\subsubsection{\textbf{Project PREDICT}}
The project PREDICT~\cite{PREDICTdataset} provides 430 datasets in 14
categories contributed by several data providers.  
From all 14 categories, just three of them are relevant to the
network intrusion detection and could be used for evaluation purposes:
\begin{compactdesc}
    \item[\textbf{Blackhole Address Space Data:}]  is collected by monitoring
    routed but unused IP address space that does not host any networked
    devices. Systems that monitor such unoccupied address space have a
    variety of names, including darkspace, darknets, network telescopes,
    blackhole monitors, sinkholes, and background radiation monitors.
    Packets observed in the darkspace can originate from a~wide range
    of security-related events, such as scanning in search of vulnerable
    targets, backscatter from spoofed denial-of-service attacks, automated spread of the Internet worms or viruses, etc. The related subcategory of this category is UCSD Archived Network Telescope Data. The archived files are in PCAP format. Source IP addresses are not anonymized.
    
    \item[\textbf{IP Packet Headers:}]  these datasets are comprised of headers
    of network data, containing information such as anonymized source
    and destination IP addresses and other IP and transport header fields.
    No payload is included. Depending on the specific dataset,
    this category of data can be used for characterization of typical
    internet traffic, or of traffic anomalies such as distributed denial
    of service attacks, port scans, or worm outbreaks.
    
    \item[\textbf{Synthetically Generated Data:}]  are generated by capturing
    information from a~synthetic environment, where benign user activity
    and malicious attacks are emulated by computer programs. In this category,
    full network packets, as well as firewall logs, application logs, and
    malicious attacks are available, without any risk of compromising
    the privacy of real people. In this category, one can know and document
    complete ``ground truth''. 
    Therefore, this category is well suited for the evaluation of NIDS systems. 
\end{compactdesc}
Note that \textbf{IDS and Firewall Data} category contains large
collection of logs submitted in a~standard format but generated from
a diverse set of hardware and software systems. It does not contain
any PCAP files, therefore it could not be used for IDS evaluation
purposes. If we were to consider categorization of datasets from the MWS project,
then mentioned datasets of PREDICT would represent probing and infection
categories.

\medskip
\subsubsection{\textbf{Project CAIDA}}
Center for Applied Internet Data Analysis (CAIDA)~\cite{caidaProject} collects several different network data types at geographically diverse locations.
The data are provided by various organizations, for whose data CAIDA guarantees anonymity and privacy. 

The CAIDA datasets are dived into three categories that reflect a status of a collection
process:
\begin{compactdesc}
\item[\textbf{Ongoing:}]  the data collection for such dataset is still
active, while collections are added regularly,
\item[\textbf{One-Time Snapshot:}]  the dataset comes from a~single collection
event that only occurred once. Future events will have a~different
dataset names,
\item[\textbf{Complete:}]  a~formerly ongoing data collection that is
finished, and will not be resumed.
\end{compactdesc}
From the network intrusion detection perspective, CAIDA
includes datasets containing e.g. DDoS attacks~\cite{caidaDDoS,caidaDDoS2},
botnet traffic~\cite{dainotti2012analysis}, dumps of various well
known worms (Conficker~\cite{caidaConficker}, Code-Red~\cite{caidaRedworm},
Witty~\cite{caidaWittyWorm}). These datasets could be utilized for the evaluation of intrusion detection approaches after a further analysis followed by labeling where it is not available. If we were to consider
a categorization of datasets from MWS project (see \autoref{sub:MWS-Datasets}), then CAIDA
would belong to probing and infection categories. 

\medskip
\subsubsection{\textbf{Project NETRESEC}}
Network Forensics and Network Security Monitoring (NETRESEC)~\cite{netresecDatasets}
is an independent software vendor with focus on the network security
field. NETRESEC specializes in software for network forensics and
analysis of network traffic. In addition, NETRESEC maintains a~comprehensive
list of publicly available PCAP files that can be used for the evaluation
of network intrusion detection approaches as well.
The datasets are divided into six categories:
\begin{compactdesc}
\item[\textbf{Cyber Defence Exercises:}] this category includes network
traffic from exercises and competitions, such as Cyber Defense Exercises
and red-team/blue-team competitions.
\item[\textbf{Capture the Flag Competitions:}] it contains files from
capture-the-flag (CTF) competitions and challenges.
\item[\textbf{Malware Traffic:}]  it contains PCAP files of captured malware
traffic from honeypots, sandboxes, and intrusions.
\item[\textbf{Network Forensics:}]  Network forensics training, challenges
and contests.
\item[\textbf{SCADA/ICS Network Captures:}] files with attacks
against Industrial Control Systems; files captured at Industrial Control
System Village (4SIC, CTF, DEF CON 22).
\item[\textbf{Uncategorized PCAP Repositories:}]  various captures that
often represent data for intrusion detection purposes.
\end{compactdesc}
If we were to consider the categorization of datasets from MWS project (see \autoref{sub:MWS-Datasets}),
then NETRESEC datasets would represent probing and infection categories. 

\medskip
\subsubsection{\textbf{DARPA 1998 and 1999 Datasets}}
The Cyber Systems and Technology Group~\cite{darpa-Url} of MIT Lincoln Laboratory has collected the first standard corpora for evaluation of network intrusion detection systems in 1998 and 1999.
There were collected two datasets DARPA 1998 and 1999, and later there were released three datasets marked as DARPA 2000, which address specific network scenarios.
If we were to consider the categorization of datasets from \autoref{sub:MWS-Datasets},
then DARPA datasets would represent probing and infection categories. 

\medskip
\subsubsection{\textbf{CCRC 2006 Dataset}}

The authors F. Massicotte et al.~\cite{massicotte2006automatic} developed a~framework for automatic evaluation of intrusion detection systems, and they collected an examplar dataset consisted of several network
attack simulations. 
We denote this dataset as CCRC 2006, because of the
main author was, at the time of the article was written, an employee of
Canada Communication Research Center in Ottawa.

The dataset is specific to signature-based network intrusion detection
systems and contains only well-known attacks, without background traffic.
The purpose of the dataset is testing and evaluation of the detection
accuracy of IDS in the case of successful and failed attack attempts.
The paper also reports an initial evaluation of the framework on two
well-known IDS, namely SORT~\cite{roesch1999snort} and Bro~\cite{paxson1999bro}. 
In the proposed framework, the authors are able to automatically generate a~large dataset, with which it is possible to automatically test and evaluate intrusion detection systems.
Note that the framework also contains a mutation layer that is able to perform various L2 and L3 protocol based obfuscations using tools Fragroute~\cite{fragroute} and Whisker~\cite{puppy1999whiskers}.
If we were to consider the categorization of datasets from \autoref{sub:MWS-Datasets},
then CCRC 2006 dataset would represent the infection category. 

\medskip
\subsubsection{\textbf{CDX 2009 Dataset}}
The CDX 2009 dataset was introduced by Sangster et al.~\cite{sangster2009toward} and it contains data in tcpdump format as well as SNORT~\cite{snortWeb} intrusion prevention logs. 
We used this dataset in our research, and it is described in \autoref{sec:desc-CDX}.
If we were to consider the categorization of datasets from the MWS project (see \autoref{sub:MWS-Datasets}),
then CDX 2009 dataset would represent probing and infection categories. 

\medskip
\subsubsection{\textbf{Twente 2009 Dataset}}
The Twente 2009 dataset~\cite{sperotto2009labeled} consists of 14.2M network flows (i.e., 155M packets) collected during a period of 9 days in 2008, where 7.6M of intrusion alerts were generated. 
The flows in this dataset were assembled by a modified version of \textit{softflowd} utility, and 98\% of them have been labeled by the authors.
The authors collected dataset by a honeypot installed on virtual host Citrix XenServer 5.
The deployed honeypot run with three opened services: OpenSSH, Apache  web server, and FTP server \textit{proftp}.

\medskip
\subsubsection{\textbf{ISCX 2012 Dataset}}
The authors of~\cite{shiravi2012toward} presented guidelines for the generation of benchmark dataset consisting of creating a malicious and benign profile that were later executed during dataset generation.
The authors generated their own dataset of network traffic (including the payload) for various network services such as HTTP, SMTP, SSH, IMAP, POP3, FTP. 
In sum, they collected 2.5M of network flows, consisting of 125M of packets.

\medskip
\subsubsection{\textbf{Contagio 2015 Dataset}}
Contagio dataset~\cite{ContagioMalwareDump} contains a collection of PCAP files from malware analysis. 
The authors collected almost 1000 malicious PCAPs from various public sources. 
The collection is irregularly updated with new PCAP files. PCAPs in the Contagio dataset
include implicit expert knowledge about the occurrence of attacks/malware. 
If we were to consider the categorization of datasets from the MWS project (see \autoref{sub:MWS-Datasets}), then the Contagio dataset would belong to categories representing an infection and malware activities after an infection.

\subsection{Datasets Consisting of High-Level Features}\label{SOA:DatasetsHighLevelFeatures}

The current category of datasets contains
representatives that were built from network traffic traces, hence it can be interpreted as a post-processed version of the former category. 
The current category of datasets contains five representatives:
KDD Cup~'99~\cite{KDDCup99web}, NSL KDD~'99~\cite{tavallaee2009NSL_KDD99}, Moore's 2005~\cite{moore2005discriminators}, Kyoto 2006+~\cite{song2011statistical}, and OptiFilter 2014~\cite{salem2014persistent}.

\medskip
\subsubsection{\textbf{KDD Cup '99}}

In 1999, KDD Cup '99~\cite{KDDCup99web} dataset was created, and it
is based on the DARPA 1998 dataset of network dumps. It has been used
for evaluating intrusion detection methods that analyze
features extracted from network traffic and host machine data. The
training dataset consists of approximately 4,9M single connection
samples from seven weeks of network traffic, each labeled as either
normal or attack, containing 41 features per connection sample. 
Similarly, the two weeks of testing data yielded around two million connection
samples. The datasets contain a~total number of 24 training attack
types, with additional 14 types in the testing dataset.
The simulated attacks fall into four main categories~\cite{KDDCup99web,tavallaee2009NSL_KDD99}: 
\begin{compactdesc}
    \item \textbf{Denial of Service Attack (DOS)}: is an attack in which the
    attacker makes some computing or memory resource too busy or too full
    to handle legitimate requests, or denies legitimate users access to
    a machine, e.g., SYN flood.

    \item \textbf{Remote to Local Attack (R2L}): occurs when an attacker who
    has the ability to send packets to a~machine over a~network, but who
    does not have an account on that machine, exploits some vulnerability
    to gain local access as a~user of that machine, e.g., guessing password,
    remote buffer overflow attacks.

    \item \textbf{User to Root Attack (U2R}): is a~class of attacks where the
    attacker begins with access to a~normal user account on the system
    (e.g., a~dictionary attack) and then is able to exploit some vulnerability to gain
    superuser access to the system, e.g. local buffer overflow
    attacks.

    \item \textbf{Probing}: is an attempt to gather information about a~network
    of computers for the purpose of circumventing its security controls,
    e.g. port scanning for vulnerable services.
\end{compactdesc}
The features of the KDD '99 dataset are, according to~\cite{KDDCup99web},
divided into three categories: 
\begin{itemize}
    \item \textbf{Basic Features:} of individual communications. This category
    encapsulates all the attributes that can be extracted from TCP or UDP communications. 

    \item \textbf{Content Features:} are extracted within a~connection suggested by domain
    knowledge. Unlike most of the DoS and Probing attacks, the R2L and
    U2R attacks cannot be described by any volumetric or frequency
    pattern. This is because the DoS and Probing attacks involve many
    connections to some hosts in a~very short period of time, while
    the R2L and U2R attacks are embedded in the data portions of the packets
    associated with a single connection. 
    To detect such attacks, features that inspect application-level behavior are employed, e.g., the number of failed login attempts. These features parse the payload of packets regardless of it is encrypted or not. 
    Hence, they cannot be extracted from only network data.

    \item \textbf{Traffic Features:} (a.k.a., time-based features) calculate statistics related to protocol behavior,
    service, etc., and they are computed using a~two-second time window.
    This category of features is further divided into two subcategories~\cite{stolfo2000cost}: 
    \begin{itemize}
        \item \textbf{Same Host Features:}  examine only the connections in the past two seconds that have the same destination host as the current connection. 
        
        \item \textbf{Same Service Features:} examine only the connections in the past two seconds that have the same service as the current connection. 
    \end{itemize}
\end{itemize}
Stolfo et al.~\cite{stolfo2000cost} criticize time-based features since there exist several slow probing attacks that scan host using a~much larger
time interval than two seconds. 
Rather than using a~time window of two
seconds, Stolfo et al.~\cite{stolfo2000cost} use a window
of 100 connections, and constructed a~mirror set of host-based traffic
features, replacing original time-based traffic features. 

\medskip
\subsubsection{\textbf{NSL KDD '99}}

Deficiencies of the KDD Cup '99 dataset were discussed in~\cite{tavallaee2009NSL_KDD99}.
The main deficiency of original dataset relates to redundant replicated entries
(78\% in the training set and 75\% in the testing set). 
The original dataset was modified, reduced, and release as the
NSL KDD '99 dataset. The training dataset contains about 130 thousand entries and the testing one about 23 thousand. In NSL KDD '99 dataset,
all samples are sorted into the original 24 classes as well as into two classes.
Complete NSL KDD '99 dataset is available at~\cite{tavallaee2012nslLink}.

\medskip
\subsubsection{\textbf{Moore's 2005}}\label{sub:Moore's-2005}
The Moore's 2005 datasets~\cite{moore2005discriminators} are primarily intended
to aid in the assessment of network traffic classification. A number
of datasets are described; each dataset consists of a~number of objects, and each object is described by a~group of features (a.k.a., discriminators~\cite{moore2005discriminators}). 
Each object within each dataset represents a~single flow of
TCP packets between a client and a server. Features for each object consist
of processed input data by discriminators extraction, and
these features serve as the input for probabilistic classification techniques. Input data is obtained by the Network Monitor tool
designed in~\cite{moore2003architecture}. 
In contrast to previously described KDD datasets, Moore's dataset
is based purely on network traffic traces, and there is not utilized
any information from host machines during the extraction of the features. 

\medskip
\subsubsection{\textbf{Kyoto 2006+}}
J. Song et al.~\cite{song2011statistical} presented an evaluation
dataset for NIDS, which was built from the 3 years of real-network traffic (since Nov. 2006 to Aug. 2009) that was collected by various types of
honeypots. 
The total number of honeypots used for collection is 348,
including two black hole sensors with 318 unused IP addresses. 
The most of honeypots were rebooted and restored original HDD image immediately
after a~malicious packet was observed. For inspection of captured
traffic, the authors use three independent security SW: SNS7160 IDS
system~\cite{symantecNetSecurity}, Clam AntiVirus software~\cite{clamAV}, and Ashula~\cite{ashula}. 
Later on, the authors have deployed SNORT~\cite{roesch1999snort} to their infrastructure. 
The dataset contains over 50 millions of normal sessions and over 43 millions
of attack sessions. The authors regarded all traffic data captured
from their honeypots as attack data and all traffic data captured
at their legitimate mail and DNS server as normal data. Also, among
the attack sessions, there were observed over 425 thousands of sessions
that were related to unknown attacks, because they did not trigger
any IDS alerts, but they contained shellcodes detected by Ashula.

The Kyoto 2006+ dataset consists of 14 statistical features taken
from the KDD Cup '99 dataset as well as 10 additional features that can
be used for further analysis and evaluation of NIDSs. 
The authors have not used any content-based features (extracted from host data) and focused only on network traffic data.
In addition to statistical features, the authors
extracted other 10 features that enable them to investigate
more effectively what kinds of attacks occurred (e.g., reflecting granularity of ground truth).
The Kyoto 2006+ dataset is available at~\cite{Kyoto2006Dataset}.

\medskip
\subsubsection{\textbf{OptiFilter 2014 -- Persistent Dataset Generation}}
Salem et al. proposed OptiFilter~\cite{salem2014persistent}, a~framework that on-the-fly constructs connection vectors from data flows. The framework collects network packets and host events continuously in real-time, parses them to a~queue of dynamic
windows, and then it generates connection vectors. Datasets
generated by the framework can be utilized for the evaluation of NIDSs.

OptiFilter handles ARP, ICMP, IP/TCP, and IP/UDP protocols. Moreover,
it utilizes a~finite state machine on TCP and UDP connections for
monitoring of their state until a~connection is closed or a~certain condition
is satisfied. 
All host-based features are collected using SNMP traps, a~mechanism
that allows systems to send messages to a~trap receiver. 
Within Windows machines, the Windows Management Instrumentation is used to filter events and send them as SNMP traps via WMI SNMP-Provider. 
In contrast, the Linux systems use syslog daemon to generate SNMP traps using the NetSNMP agent. 
The extracted features of OptiFilter framework are influenced by KDD
Cup 99~\cite{KDDCup99web} and Kyoto 2006~\cite{song2011statistical}
datasets and consist of three categories:
\begin{compactdesc}
    \item \textbf{Network-based Features:} timestamp, source and destination
    IPs, ports, protocol type, service, transferred Bytes,
    connection state (using BRO~\cite{paxson1999bro}),
    the number of fragmentation errors.
    
    \item \textbf{Traffic Features:} are statistical and are
    derived from the basic features. They are divided into two types,
    time-based traffic features, and connection-based traffic features.
    Both types are distinguished and treated differently. The former type is calculated based on a~dynamic time window, e.g., the last five seconds, while the latter type is calculated on a~configurable
    connection window, e.g., the last 1000 connections. 
    
    \item \textbf{Content Features:} the features are obtained directly from a monitored host using SNMP. Examples are the number of failed login attempts, the indication of a successful login, and the indication of obtaining a root shell.
\end{compactdesc}
\smallskip\noindent
In the evaluation, the authors generated a dataset called SecMonet, in which 17 common services were captured (e.g., FTP, SSH, telnet, SMTP, SMB, NFS, etc.).
However, it is not clear whether the dataset contains a self-collected malicious traffic, or it is only substituted from KDD Cup '99.
\medskip
\subsubsection{\textbf{CICIDS 2017 Dataset}}
CICIDS 2017 dataset~\cite{sharafaldin2018toward} consists of network attacks such as DoS, DDoS, Brute force, XSS, SQL injection, Heartbleed, infiltration through the scam, and port scanning. 
The authors generated benign data based on the extracted profile from an analysis of 25 users, which is in line with the approach proposed in~\cite{shiravi2012toward}.
The infrastructure used for the data collection consisted of 15 vulnerable Linux-based \& Windows-based machines and 4 attacker machines.
Further, the authors extracted 80 features using CICFlowMeter tool~\cite{lashkari2017characterization} and provided them along with the network traffic traces.

\vspace{0.3cm}
\section{Discussion}\label{sec:discussion}
\paragraph{Age of Vulnerabilities}
Although there exist a plethora of publicly available exploit-codes for contemporary vulnerabilities, the situation with corresponding available vulnerable SW is more difficult due to understandable prevention reasons imposed by SW vendors.
Therefore, we were able to contain only older available high-severity vulnerable services that are outdated.
However, we conjecture that from the point of view of non-payload-based network intrusion detection (not inspecting the payload of packets), the behavioral characteristics of simulated high-severity attacks are similar regardless of the age of vulnerabilities.
In particular, we refer to the buffer overflow attacks, which are executed in a few stages involving a repeated transfer of one or more packets with the maximum payload.

\paragraph{Cross-Dataset Evaluation}
In this paper, we provided only a basic benchmarking of several supervised classifiers on ASNM datasets.
Nevertheless, it is worth to note that different benchmarking techniques can be used as well.
One example is cross-dataset evaluation, where the target classifier is trained on the input data of one dataset, and then it is evaluated on data taken from another dataset.
We leave these tasks as an open challenge for the community. 
\vspace{0.3cm}
\section{Conclusion}\label{sec:conclusion}
In this paper, we presented three datasets consisting of extracted high-level network features (ASNM features).
These datasets are intended for non-payload-based network intrusion detection and adversarial classification, enabling to test evasion resistance of machine learning-based classifiers.
In detail, ASNM-CDX-2009 dataset might serve for basic benchmarking of machine learning-based classifiers, while ASNM-TUN and ASNM-NPBO datasets might serve for more advanced benchmarking of these classifiers, such as testing the classifiers on evasion resistance.
In future work, we will extend ASNM datasets with data collected from other experiments.
 
\vspace{0.5cm}
\bibliographystyle{IEEEtran}

\bibliography{ref}

\vspace{0.3cm}
\appendix 
\label{sec:appendix}
\subsection{Tuples of TCP Connection and Packet}
We detail the TCP connection tuple in \autoref{tab:TCP_connection_tuple_items}.
In the table, the superscript at $\mathbb{R}^*$ represents a set of positive real numbers, including zero. 
Next, we present a detailed description of items from the packet tuple in \autoref{tab:Packet_tuple_items}.
The description contains assignment to a particular layer of ISO/OSI stacked model together with supported instances of the protocols -- placeholder $*$ represents an arbitrary protocol instance. 
Also note that in the case of \textit{data} field, superscript in $X^*$ represents an iteration of the set $X$.

	\begin{table}
	    \centering
	    \begin{tabular}{r r l}
	       	\toprule
	        \textbf{Symbol} & &\textbf{Description}  \\ 
			\midrule
	        $t_s \in \mathbb{R}^{+}$            & & Timestamp of the connection's start.\\
	        $t_e \in \mathbb{R}^{+}$            & & Timestamp of the connection's end.\\
	        $p_{c} \in \{0,\ldots,2^{16}-1\}$          & & Port of the client within the TCP connection.\\
	        $p_{s} \in \{0,\ldots,2^{16}-1\}$          & & Port of the server within the TCP connection.\\
	        $ip_{c} \in \{0,\ldots,2^{32}-1\}$         & & IPv4 address of the client.\\
	        $ip_{s} \in \{0,\ldots,2^{32}-1\}$         & & IPv4 address of the server.\\
	        $P_{c}$          & & Set of packets sent by client to server.\\
	        $P_{s}$          & & Set of packets sent by server to client.\\
			
			\bottomrule			

	    \end{tabular}  
    	\caption{Symbols of the TCP connection tuple.}     	    
	    \label{tab:TCP_connection_tuple_items}
	\end{table}

\newcommand{\parboxWidthI}{0.52\columnwidth}
\begin{table}
	\centering
	\begin{tabular}{@{}r l@{}}
		\toprule
		\textbf{Symbol} & \textbf{Description} \\
		\midrule
		$t \in \Re^{+}_{0}$
		& \parbox{\parboxWidthI}{\vspace{0.1cm}Relative time of the packet capture (L1, $*$).} \\
		$size \in \aleph $
		& \parbox{\parboxWidthI}{\vspace{0.1cm}Size of the whole Ethernet frame including Ethernet header (L1, $*$).} \\
		$eth_{src} \in \{0,\ldots,2^{48}-1\}$
		& \parbox{\parboxWidthI}{\vspace{0.1cm}Source MAC address of the frame  (L2, Ethernet).} \\
		$eth_{dst} \in \{0,\ldots,2^{48}-1\}$ 
		& \parbox{\parboxWidthI}{\vspace{0.1cm}Destination MAC address of the frame (L2, Ethernet).} \\
		$ip_{off} \in \{0,\ldots, 2^{13}-1\}$
		& \parbox{\parboxWidthI}{\vspace{0.1cm}Offset field (L3, IPv4).} \\
		$ip_{ttl} \in \{0,\ldots, 2^{8}-1\}$
		& \parbox{\parboxWidthI}{\vspace{0.1cm}Time to live field (L3, IPv4).} \\
		$ip_p \in \{0,\ldots, 2^{8}-1\}$
		& \parbox{\parboxWidthI}{\vspace{0.1cm}Protocol field (L3, IPv4).} \\
		$ip_{sum} \in \{0,\ldots, 2^{16}-1\}$
		& \parbox{\parboxWidthI}{\vspace{0.1cm}Checksum of the header (L3, IPv4).} \\
		$ip_{src} \in \{0,\ldots, 2^{32}-1\}$
		& \parbox{\parboxWidthI}{\vspace{0.1cm}Source IP address of the packet (L3, IPv4).} \\
		$ip_{dst} \in \{0,\ldots, 2^{32}-1\}$
		& \parbox{\parboxWidthI}{\vspace{0.1cm}Destination IP address of the packet (L3, IPv4).} \\
		$ip_{dscp} \in \{0,\ldots, 2^{8}-1\}$
		& \parbox{\parboxWidthI}{\vspace{0.1cm}Differentiated services code point field (L3, IPv4).} \\
		$tcp_{sport} \in \{0,\ldots, 2^{16}-1\}$
		& \parbox{\parboxWidthI}{\vspace{0.1cm}Source port of the packet (L4, TCP).} \\
		$tcp_{dport} \in \{0,\ldots, 2^{16}-1\}$
		& \parbox{\parboxWidthI}{\vspace{0.1cm}Destination port of the packet (L4, TCP).} \\
		$tcp_{sum} \in \{0,\ldots, 2^{16}-1\}$
		& \parbox{\parboxWidthI}{\vspace{0.1cm}Checksum of the header (L4, TCP).} \\
		$tcp_{seq} \in \{0,\ldots, 2^{32}-1\}$
		& \parbox{\parboxWidthI}{\vspace{0.1cm}Sequence number of the packet (L4, TCP).} \\
		$tcp_{ack} \in \{0,\ldots, 2^{32}-1\}$
		& \parbox{\parboxWidthI}{\vspace{0.1cm}Acknowledgment number of the packet (L4, TCP).} \\
		$tcp_{off} \in \{0,\ldots, 2^{8}-1\}$
		& \parbox{\parboxWidthI}{\vspace{0.1cm}Offset and reserved fields together (L4, TCP).} \\
		$tcp_{flags} \in \{0,\ldots, 2^{8}-1\}$ 
		& \parbox{\parboxWidthI}{\vspace{0.1cm}Control bits (L4, TCP).} \\
		$tcp_{win} \in \{0,\ldots, 2^{16}-1\}$
		& \parbox{\parboxWidthI}{\vspace{0.1cm}Window field (L4, TCP).} \\
		$tcp_{urp} \in \{0,\ldots, 2^{16}-1\}$
		& \parbox{\parboxWidthI}{\vspace{0.1cm}Urgent pointer field (L4, TCP).} \\
		$data \in \{0,\ldots,2^{8}-1\}^{\ast}$
		& \parbox{\parboxWidthI}{\vspace{0.1cm}Payload of the packet (L7, $*$).} \\
		\bottomrule
	\end{tabular}     
	\caption{Symbols of the packet tuple.}
	\label{tab:Packet_tuple_items} 
\end{table} \hspace{-0.5cm}

\subsection{FFS Selected Features}
As part of benchmarking ASNM datasets, in this section, we enumerate subsets of the ASNM features selected by the FFS method with Naive Bayes classifier running over 5-fold cross-validation.
We present ASNM features selected using: (1) the ASNM-CDX-2009 dataset in \autoref{FFS-features-CDX},  (2) the ASNM-TUN dataset in \autoref{FFS-features-tunneling}, and (3) the ASNM-NPBO dataset in \autoref{FFS-features-NPBO}.
Note that the FFS DL set denotes features selected when only legitimate samples and direct attacks were included in the FFS experiment.
The FFS DOL set denotes selected features when, in addition to the previous case, obfuscated attacks were included in the FFS experiment.

\begin{table}
	\centering    
	\begin{tabular}{@{}l p{0.05cm} p{5.0cm} @{}}
		\toprule
		
		\textbf{Feature ID} & & \textbf{Description} \\                   
		\midrule

		\textbf{PolyInd3ordOut[0]}  & & $\bullet$  Approximation of outbound communication (from client to server) by polynomial of 3rd order in the index domain of packet occurrences.
		The feature represents the 1st coefficient of the approximation,\\
		\textbf{PolyInd3ordOut[3]}  & & $\bullet$  The same as the previous one, but the feature represents the 4th coefficient of the approximation,\\ 
		\textbf{PolyInd8ordOut[6]}  & & $\bullet$  The same as the previous ones, but the feature represents the 7th coefficient of the approximation,\\ 
		\textbf{InPkt1s10i[7]}  & & $\bullet$  Lengths of inbound packets occurred in the first second of a~connection which are distributed into 10 intervals. 
		The feature represents totaled inbound packet lengths of the 8th interval,\\ 
		\textbf{InPkt1s10i[0]}  & & $\bullet$  The same as the previous one, but it represents the 1st interval, \\ 
		\textbf{InPkt1s10i[1]}  & & $\bullet$  The same as the previous one, but it represents the 2nd interval,\\ 
		\textbf{GaussProds8All[7]}  & & $\bullet$  Normalized products of all packet sizes with 8 Gaussian curves. 
		The feature represents a~product of the 8th slice of packets with a~Gaussian function which fits the interval of the packets' slice. \\

		\bottomrule            
	\end{tabular}       
	\caption{ASNM features selected by FFS with the Naive Bayes classifier using ASNM-CDX-2009 dataset.}
	\label{FFS-features-CDX}
\end{table}

\begin{table}
	\centering    
	\begin{tabular}{@{}l p{0.05cm} p{5.0cm} c r c@{}}
		\toprule
		
		\textbf{Feature ID} & & \multicolumn{1}{c}{\textbf{Description}} & & \begin{turn}{90} \textbf{FFS DOL} \end{turn} & \begin{turn}{90} \textbf{FFS DL} \end{turn} \\                   
		\midrule        
		
		SigPktLenIn   &    & $\bullet$ Std. deviation of inbound (server to client) packet sizes. & & X &  \\        
		
		ConTcpFinCntIn &    & $\bullet$ The number of TCP FIN flags occurred in inbound traffic. & & X & X \\
		
		ConTcpSynCntIn &    & $\bullet$ The number of TCP SYN flags occurred in inbound traffic. & & X & X \\
		
		InPktLen32s10i[0]   &    & $\bullet$ Lengths of inbound packets occurred in the first 32 seconds of a~connection which are distributed into 10 intervals. The feature represents totaled inbound packet lengths of the 1st interval. & & X & \\    
		
		InPktLen1s10i[2]   &    & $\bullet$ The same as the previous one, but computed above the first second of a~connection.  The feature represents totaled inbound packet lengths of the 3rd interval.  & & X & \\
		
		InPktLen8s10i[7]   &    & $\bullet$ The same as the previous one, but computed above the first 8 seconds of a~connection.  The feature represents totaled inbound packet lengths of the 8th interval.  & & X & \\
		
		OutPktLen1s10i[0]   &  & $\bullet$ Lengths of outbound (client to server) packets occurred in the first second of a~connection which are distributed into 10 intervals. The feature represents totaled outbound packet lengths of the 1st interval. & & X &  \\    
		
		FourGonAngleN[9]$^*$   &    & $\bullet$ Fast Fourier Transformation (FFT) of all packet sizes. The feature represents the angle of the 10th coefficient of the FFT in goniometric representation. & & X & X \\
		
		InPktLen8s10i[1]   &    & $\bullet$ Lengths of inbound packets occurred in the first 8 seconds of a~connection which are distributed into 10 intervals. The feature represents totaled inbound packet lengths of the 2nd interval. & &  & X\\    
		
		PolyInd8ordOut[5]   &    & $\bullet$ Approximation of outbound packet lengths in index domain by polynomial of 8th order. The feature represents 6th coefficient of the polynomial.  & &  & X\\  
		
		PolyInd8ordIn[5]   &    & $\bullet$ Approximation of inbound packet lengths in index domain by polynomial of 8th order. The feature represents 6th coefficient of the polynomial.  & &  & X\\

		\midrule
		\multicolumn{6}{p{7.0cm}}{$^*$Sizes of inbound and outbound packets are represented by negative and positive values, respectively.}\\
		
		\bottomrule            
	\end{tabular}       
	\caption{ASNM features selected by FFS with the Naive Bayes classifier using ASNM-TUN dataset.}
	\label{FFS-features-tunneling}
\end{table}

\setlength{\tabcolsep}{2.0pt}
\begin{table}
	\centering
	\footnotesize{            
		\begin{tabular}{@{}l p{0.02cm} p{5.2cm} c c c c  @{}}
			\toprule
			
			\textbf{Feature ID} & & \multicolumn{1}{c}{\textbf{Description}} & & \begin{turn}{90} \textbf{FFS DOL} \end{turn} & \begin{turn}{90} \textbf{FFS DL} \end{turn} \\                   
			
			\midrule

			SigPktLenOut   &    & $\bullet$ Std. deviation of outbound (client to server) packet sizes. & & X & X  \\    
			
			MeanPktLenIn   &    & $\bullet$ Mean of packet sizes in inbound traffic of a connection. & & X & X \\
			
			CntOfOldFlows &    & $\bullet$ The number of mutual connections between client and server, which started up to 5 minutes before start of an analyzed connection. & & X & X \\
			
			CntOfNewFlows &    & $\bullet$ The number of mutual connections between client and server, which started up to 5 minutes after the end of an analyzed connection. & & X & X \\
			
			ModTCPHdrLen   &     & $\bullet$ Modus of TCP header lengths in all traffic.  & & X & \\
			
			UrgCntIn &    & $\bullet$ The number of TCP URG flags occurred in inbound traffic. & & X & \\
			
			FinCntIn &    & $\bullet$ The number of TCP FIN flags occurred in inbound traffic. & & & X  \\
			
			PshCntIn &    & $\bullet$ The number of TCP PUSH flags occurred in inbound traffic. & & & X \\
			
			FourGonModulIn[1]   &    & $\bullet$ Fast Fourier Transformation (FFT) of inbound packet sizes. The feature represents the module of the 2nd coefficient of the FFT in goniometric representation. & & X & X \\
			
			FourGonModulOut[1]   &    & $\bullet$ The same as the previous one, but it represents the module of the 2nd coefficient of the FFT for outbound traffic. & & & X \\
			
			FourGonAngleOut[1]   &    & $\bullet$ The same as the previous one, but it represents the angle of the 2nd coefficient of the FFT. & & X &  \\
			
			FourGonAngleN[9]   &    & $\bullet$ Fast Fourier Transformation (FFT) of all packet sizes, where inbound and outbound packets are represented by negative and positive values, respectively. The feature represents the angle of the 10th coefficient of the FFT in goniometric representation. & & X & X \\    
			
			FourGonAngleN[1]   &    & $\bullet$ The same as the previous one, but it represents the angle of the 2nd coefficient of the FFT. & & X & \\
			
			FourGonModulN[0]   &    & $\bullet$ The same as the previous one, but it represents the module of the 1st coefficient of the FFT. & & & X \\
			
			PolyInd13ordOut[13]   &    & $\bullet$ Approximation of outbound communication by polynomial of 13th order in the index domain of packet occurrences. The feature represents the 14th coefficient of the approximation.  & & X & \\
			
			PolyInd3ordOut[3]   &    & $\bullet$ The same as the previous one, but it represents the 4th coefficient of the approximation.  & & & X \\
			
			GaussProds8All[1] &    & $\bullet$ Normalized products of all packet sizes with 8 Gaussian curves. The feature represents a product of the 2nd slice of packets with a Gaussian function that fits the interval of the packets' slice.   & & X &  \\                
			
			GaussProds8Out[7] &    & $\bullet$ The same as the previous one, but computed above outbound packets and represents a product of the 8th slice of packets with a Gaussian function that fits the interval of the packets' slice. & & & X \\  
			
			InPktLen1s10i[5]   &    & $\bullet$ Lengths of inbound packets occurred in the first second of a connection, which are distributed into 10 intervals. The feature represents totaled outbound packet lengths of the 6th interval. & & X & \\
			
			OutPktLen32s10i[3] &    & $\bullet$ The same as the previous one, but computed above the first 32 seconds of a connection.  The feature represents totaled outbound packet lengths of the 4th interval.  & &  & X \\
			
			OutPktLen4s10i[2] &    & $\bullet$ The same as the previous one, but computed above the first 4 seconds of a connection.  The feature represents totaled outbound packet lengths of the 3rd interval.  & &  & X \\
			
			\bottomrule    
			
		\end{tabular}
	}
	\caption{ASNM features selected by FFS using the Naive Bayes classifier.}
	\label{FFS-features-NPBO}
\end{table}
 
\end{document}